\documentclass[twocolumn,showpacs,nofootinbib,preprintnumbers,amsmath,amssymb,showkeys,superscriptaddress]{revtex4-1}
\usepackage{graphicx}
\usepackage{amsmath}
\usepackage{amssymb}
\usepackage{dcolumn}
\usepackage{bm}
\usepackage{xcolor}
\usepackage{dsfont}
\usepackage{feynmp}
\usepackage{marvosym}
\usepackage{phaistos}
\usepackage[hidelinks=true]{hyperref}

\usepackage{ulem}

\newcommand \beq{\begin{eqnarray}}
\newcommand \eeq{\end{eqnarray}}

\begin{document}
\unitlength=1mm
\allowdisplaybreaks

\title{The center-symmetric Landau gauge meets the lattice}

\author{Duifje Maria van Egmond}
\affiliation{ICTP South American Institute for Fundamental Research Instituto de F\'isica Te\'orica, UNESP - Univ. Estadual Paulista Rua Dr. Bento Teobaldo Ferraz 271, 01140-070, S\~ao Paulo, SP, Brazil.}

\author{Orlando Oliveira}
\affiliation{CFisUC, Department of Physics, University of Coimbra, P-3004 516 Coimbra, Portugal.}

\author{Urko Reinosa}
\affiliation{Centre de Physique Th\'eorique, CNRS, Ecole polytechnique, IP Paris, F-91128 Palaiseau, France.}

\author{Julien Serreau}
\affiliation{Universit\'e Paris Cit\'e, CNRS, Astroparticule et Cosmologie, F-75013 Paris, France.}

\author{Paulo J. Silva}
\affiliation{CFisUC, Department of Physics, University of Coimbra, P-3004 516 Coimbra, Portugal.}

\author{Matthieu Tissier}
\affiliation{Sorbonne Universit\'e, CNRS, Laboratoire de Physique Th\'eorique de la 
Mati\`ere Condens\'ee, 75005 Paris, France.}

\date{\today}

\begin{abstract}
A lattice implementation of the recently introduced center-symmetric Landau gauge is discussed and its predictions confronted with numerical Monte Carlo simulations. It is shown that the link average and the link correlators computed in that gauge are order parameters of the confinement-deconfinement transition at nonzero temperature. Strictly speaking, this requires a specific treatment of the Gribov copies that we discuss in detail. The numerical simulations comply with the theoretical predictions for the link average computed below and above the deconfinement temperature. Our results show that, within appropriately chosen gauges, one can construct local order parameters for center symmetry, as proxies for the non-local Polyakov loop.
\end{abstract}

\maketitle

\tableofcontents

\section{Introduction}
Center symmetry plays a pivotal role in the context of non-Abelian gauge theories and in particular Quantum Chromodynamics (QCD). Although not a symmetry of the QCD action, it becomes a strict symmetry in the limit of infinitely heavy quarks \cite{Pisarski:2002ji,Greensite:2011zz}, just as chiral symmetry becomes a strict symmetry in the limit of massless quarks. And just as chiral symmetry is broken spontaneously at low temperatures, providing a mechanism of mass generation for the confined hadronic bound states, center symmetry can be broken at high temperatures, leading to a quark-gluon plasma phase \cite{Gavai:1982er,Gavai:1983av,Celik:1983wz,Svetitsky:1985ye}. It is then no doubt that these two symmetries play such a prominent role in the investigation of the phase structure of strongly interacting nuclear matter.

Describing the Yang-Mills confinement/deconfinement transition in the continuum is not straightforward, however. One issue comes from the necessary gauge fixing that can break the underlying center symmetry explicitly at the level of the gauge-fixed action \cite{vanEgmond:2023lnw}. Although this breaking should not be visible at the level of gauge-invariant order parameters such as the Polyakov loop \cite{Polyakov:1978vu}, the latter are generically nonlocal observables, which are not of easy access in continuum calculations \cite{Herbst:2015ona}. Moreover, gauge-invariant quantities can also be contaminated by the gauge-fixing symmetry breaking due to the necessary approximations or modeling employed in such approches. Of course, this is a spurious effect which should diminish as one improves the degree of approximation, but it can have harmful consequences at the lowest orders of approximation.

For these various reasons, in the continuum, it is interesting to use gauge fixings that do not break the center symmetry explicitly, while offering the possibility to define alternative (and local) order parameters which are simpler to evaluate in practice. A major step in this direction was achieved in Refs.~\cite{Braun:2007bx,Braun:2010cy} where it was shown that the gauge-field average plays the role of an order parameter for center symmetry in the self-consistent background Landau gauge \cite{Abbott:1980hw,Abbott:1981ke}. This approach, which has been quite successful in recent years \cite{Fister:2013bh}, is not void of subtleties, however. In particular, the gauge-field average is obtained from the minimization of a functional which is not exactly a thermodynamical potential in the sense of a Legendre transform of the free-energy. One consequence is that the very identification of the gauge-field average with the minimum of this potential relies on additional assumptions which are not necessarily fulfilled in the presence of approximations or modeling \cite{Reinosa:2020mnx}.

Recently, an alternative proposal has been put forward in \cite{vanEgmond:2021jyx,vanEgmond:2023lnw} where, rather than working with a self-consistent background, one uses a fixed center-symmetric background configuration both in the confined and deconfined phases. It was shown that in this so-called center-symmetric Landau gauge, the gauge field average, and more generally, the gauge-field correlators are, in fact, local order parameters for center symmetry \cite{vanEgmond:2023lfu}. The main advantage here is that the functional to be extremized is a genuine Legendre transform, which thus eliminates the above-mentioned subtleties of the self-consistent background field approach.

Various tests of these theoretical expectations have been performed, showing that, in the center-symmetric Landau gauge, the one-point function \cite{vanEgmond:2021jyx} but also the two-point correlators \cite{vanEgmond:2022nuo} capture properly the nature of the transition as a function of the gauge group and give very good results for the deconfinement transition. So far, these tests have been conducted within the context of the Curci-Ferrari model \cite{Curci:1976bt}, a simple extension of the FP gauge fixed Lagrangian which successfully captures many infrared aspects of YM theories and which offers the great advantage of allowing for  simple (semi)analytical calculations (see Ref.~\cite{Pelaez:2021tpq} for a recent review).

Although the Curci-Ferrari model was pivotal in some of the above developments as it provided a framework where many quantities could be computed semi-analytically, a serious test of the relevance of the center-symmetric Landau gauge should lift any source of modeling. In this work, we then propose a lattice implementation of the center-symmetric gauge fixing. We shall here restrict to the SU($3$) case but the discussion can be easily extended to SU(N) along the lines of Ref.~\cite{vanEgmond:2023lfu}.

In Sect.~\ref{sec:gen}, we gather some generalities regarding the lattice set-up, including the notions of links and plaquettes, the Wilson action and the way observables are computed using Monte-Carlo importance sampling. Section \ref{sec:center} discusses center symmetry and its breaking from the point of view of the gauge-invariant Polyakov loop. We put special emphasis on the very different nature of the Monte-Carlo ensemble in the broken and unbroken phases. We also discuss charge conjugation as it plays a role in the following. In Sect.~\ref{sec:links}, we repeat the same discussion from the point of view of link correlators, which, in a sense, replace the gauge-field correlators of the continuum approach. As the correlators are gauge-dependent quantities, defined only once a gauge is fixed, we spend some time recalling the lattice gauge fixing procedure, paying particular attention to the issue of Gribov copies. We investigate the general conditions under which the link correlators actually become order parameters for center symmetry. This defines the class of center-symmetric gauges. Sect.~\ref{sec:cslg} considers an explicit example of a center-symmetric gauge, closely related to the continuum one proposed in Ref.~\cite{vanEgmond:2021jyx}. We identify and solve the symmetry constraints for the link average in that gauge and study whether and when these constraints are fulfilled within actual numerical simulations. Finally, Sect.~\ref{sec:add} gathers additional material such as an equivalent analysis in terms of so-called twisted link variables, some remarks concerning the relation between the links and and the gauge field, or the invariance properties of the gauge-fixed ensemble.

As a final word, let us mention that there exist various lattice studies that aim at tracking down the transition using gauge-field correlators in non-center-symmetric gauges, such as the standard Landau gauge \cite{Cucchieri:2010lgu,Fischer:2010fx,Maas:2011ez,Mendes:2015jea,Silva:2013maa,Aouane:2011fv}, alongside similar studies in the continuum \cite{Fister:2011uw,Quandt:2015aaa}. Although manifest changes of behavior have been reported for these quantities in the vicinity of the transition, some of which seem to mimic those of an actual order parameter, no rigorous connection exists to date between those observed behaviors and center-symmetry breaking. It is the purpose of the present work, together with Refs.~\cite{vanEgmond:2021jyx,vanEgmond:2023lfu}, to provide a setup where such a connection can be clearly established.

One important exception is the study of Ref.~\cite{Silva:2016onh} where a connection between center-symmetry breaking and differences of Landau gauge propagators over different ensembles of configurations (corresponding to different center sectors) was established. We stress that these results are not in contradiction with the discussion in the present work, but rather complementary. A detailed account of the two approaches will be discussed in a forthcoming work dedicated to the propagators.

\section{Generalities}\label{sec:gen}
We consider the lattice discretization of SU(3) Yang-Mills theory using link and plaquette variables. The sites of the lattice are described by a $4$-vector $n$ with integer components $\smash{n_\mu\in\mathds{N}}$, with $\smash{\mu=1,2,3,4}$, $\smash{0\leq n_\mu\leq L_\mu-1}$ and where $L_\mu$ is the lattice extent along direction $\mu$.  

\subsection{Links and plaquettes}
We denote by $\hat\mu$ the vector of components $\smash{\hat\mu_\nu\equiv\delta_{\mu\nu}}$ which connects a given site $n$ to its neighboring site $n+\hat\mu$ along direction $\mu$. To such an oriented link, one associates a link variable $\smash{U_\mu(n)\in}$ SU(3). Correspondingly, the oriented link connecting $\smash{n+\hat\mu}$ to $n$ is associated with the link variable $U_\mu^\dagger(n)$ also written as $U_{-\mu}(n+\hat\mu)$. The thermal field theory set-up considered in this work assumes periodic boundary conditions in all directions. The link variables can thus be extended over an infinite lattice such that $\smash{U_\mu(n+L_\nu\hat\nu)=U_\mu(n)}$ for any direction $\nu$.

With the links at our disposal, we can consider more general constructs. In particular, an oriented plaquette is a square loop made of oriented links, starting at $n$ and going through $n+\hat\mu$, $n+\hat\mu+\hat\nu$, $n+\hat\nu$ and back to $n$. To each such oriented loop, one associates the plaquette variable
\beq
U_{\mu\nu}(n) & \equiv & U_\mu(n)U_\nu(n+\hat\mu)U_{-\mu}(n+\hat\mu+\hat\nu)U_{-\nu}(n+\hat\nu)\nonumber\\
& = & U_\mu(n)U_\nu(n+\hat\mu)U^\dagger_{\mu}(n+\hat\nu)U^\dagger_{\nu}(n)\,,
\eeq
which is also an element of SU(3).

\subsection{Wilson action and gauge invariance}
The plaquette variables enter the definition of the Wilson action
\beq
S_W[U]\equiv\beta\sum_n\sum_{\mu<\nu} {\rm Re}\,{\rm tr}\,\big[\mathds{1}-U_{\mu\nu}(n)\big]\,,\label{eq:SW}
\eeq
where $\smash{\beta = 2N/g^2}$ and $g$ denotes the bare coupling constant. The Wilson action is a discretized version of the gauge-invariant Yang-Mills action. Let us briefly recall how gauge invariance comes about.

A gauge transformation is defined by the choice, at each site $n$ of the lattice, of an element $\smash{g_0(n)\in}$ SU(3) such that
\beq
g_0(n+L_\nu\hat\nu)=g_0(n)\,,\label{eq:periodic}
\eeq 
for any direction $\nu$ (periodicity in all directions). Gauge transformations form a group denoted ${\cal G}_0$ in what follows. This includes global color rotations.
Under a gauge transformation
\beq
U_\mu(n)\to g_0(n)U_\mu(n)g_0^\dagger(n+\hat\mu)\equiv U^{g_0}_\mu(n)\,,\label{eq:linkg}
\eeq
the plaquette variables transform as
\beq
U_{\mu\nu}(n)\to g_0(n)U_{\mu\nu}(n)g_0^\dagger(n)\,,
\eeq
and the Wilson action (\ref{eq:SW}) is invariant.

\subsection{Observables and Monte-Carlo ensemble}
Observables are obtained as expectation values of gauge-invariant functionals (that is, such that $\smash{{\cal O}[U^{g_0}]={\cal O}[U]}$) weighted by the gauge-invariant probability distribution $\exp\{-S_W[U]\}$:
\beq
\langle {\cal O}\rangle\equiv \frac{\int \prod_{n,\mu}dU_\mu(n)\,{\cal O}[U]\,e^{-S_W[U]}}{\int \prod_{n,\mu}dU_\mu(n)\,e^{-S_W[U]}}\,.
\eeq
In practice, this large multi-dimensional integral is evaluated using importance sampling Monte-Carlo techniques. One first generates a finite ensemble $\smash{{\cal E}=\{U_\mu(n)\}}$ of link configurations that follow the probability distribution $\exp\{-S_W[U]\}$. Then, the observable is evaluated as
\beq
\langle {\cal O}\rangle\simeq\frac{1}{N_{\rm conf}}\sum_{U_\mu\in{\cal E}}{\cal O}[U]\equiv \langle {\cal O}\rangle_{\cal E}\,.
\eeq
We shall refer to ${\cal E}$ as a Monte-Carlo ensemble.

We note that the gauge-invariance of $S_W[U]$ reflects the redundancy of the description of the system in terms of gauge fields and the Monte-Carlo ensemble generated from $S_W[U]$ should reflect this redundancy. That is, for a sufficiently large ensemble, for any link configuration $U_\mu(n)$ in the ensemble, its ${\cal G}_0$-orbit, defined as $\{U^{g_0}_\mu\,|\,g_0\in {\cal G}_0\}$, is faithfully represented in the ensemble. Of course, in practice, the gauge ensembles are finite, not necessarily large, but should provide good representations of the gauge manifold.

\section{Center symmetry}\label{sec:center}
The gauge transformations in ${\cal G}_0$ leave the Wilson action invariant while preserving the periodicity of the link variables. As we now recall, however, these are not the only transformations endowed with these properties.
 
\subsection{Center transformations}
In particular, one can consider transformations
\beq
U_\mu(n)\to  g(n)U_\mu^\dagger(n)g^\dagger(n+\hat\mu)\equiv U_\mu^g(n)\,,\label{eq:V_transfo}
\eeq
that are periodic in the time direction but only modulo an element of the center of SU($3$):\footnote{We could also consider center transformations along the other Euclidean directions. However, in the present finite temperature context, one should eventually consider the limit $L_\nu\to\infty$ with a fixed inverse temperature $\smash{L_4a=1/T}$ and $\smash{L_i a\to \infty}$, where $a$ denotes the lattice spacing. In this case, only the temporal center transformations are relevant.}
 \beq
g(n+L_4\hat{4})=e^{i2\pi/3}g(n)\,.\label{eq:modulo}
\eeq
These are known as center transformations. Of course, $g^\dagger$ is also a center transformation with associated center element $e^{-i2\pi/3}$. Together with the periodic gauge transformations, they form a group denoted ${\cal G}$ in what follows, which contains ${\cal G}_0$ as a subgroup. From now on, we will exclusively use $g_0$ to denote periodic gauge transformations, i.e. transformations that belong to ${\cal G}_0$, while $g$ or $g^\dagger$ will denote center transformations with a non-trivial center-element.

That center transformations are symmetries of the Wilson action is obvious from the discussion in the previous section. However, even though they take the same form as a gauge transformations, they do not qualify as a genuine gauge transformations (in the sense of redundancies in the description of the system at hand) but, rather, as physical transformations. This is because they act non-trivially on some observables, an example of which is the Polyakov loop.

 \subsection{The Polyakov loop}
The Polyakov loop is defined as the trace of a product of links that loop around the temporal direction:
\beq
\Phi[U](\vec{n}) & \equiv & \frac{1}{3}{\rm tr}\,U_4(n_0)U_4(n_0+\hat{4})U_4(n_0+2\hat{4})\cdots\nonumber\\
& & \hspace{1.0cm}\cdots U_4(n_0+(L_4-1)\hat{4})\,,\label{eq:Phi_func}
\eeq
with $\smash{n_0\equiv (\vec{n},0)}$.\footnote{We recall that $\smash{\hat{4}=(\vec{0},1)}$ so that $\smash{n_0+p\hat{4}=(\vec{n},p)}$.} 

Under a periodic gauge transformation $g_0$, one has
\beq
\Phi[U^{g_0}](\vec{n})=\Phi[U](\vec{n})\,.\label{eq:gauge_inv}
\eeq
The Polyakov loop is then gauge-invariant and its expectation value defines an observable
\beq
\langle\Phi\rangle\equiv \frac{\int \prod_{n,\mu}dU_\mu(n)\,\Phi[U](\vec{n})\,e^{-S_W[U]}}{\int \prod_{n,\mu}dU_\mu(n)\,e^{-S_W[U]}}\,,
\eeq 
referred to in what follows as the averaged Polyakov loop. Due to the translation invariance of the Wilson action and the use of periodic boundary conditions, this average does not depend on $\vec{n}$. As is well known, its physical interpretation is that, in the infinite volume and continuum limits, it corresponds to $e^{-\Delta F/T}$ where $\Delta F$ is the free-energy of a static color charge in a thermal bath of gluons at temperature $T$. It follows that the averaged Polyakov loop can be used to probe the confinement/deconfinement transition: a vanishing averaged Polyakov loop signals a confined phase ($\smash{\Delta F=\infty}$), whereas a non-zero value signals a deconfined phase ($\smash{\Delta F<\infty}$).

That $\langle\Phi\rangle$ can behave so differently depending on the temperature relies on the fact that the symmetry (\ref{eq:V_transfo}) can either be explicitly realized or spontaneously broken. One rigorous way to discuss this is to introduce a small perturbation to the Wilson action in the form of a source term\footnote{The source $\rho e^{i\theta}$ coupled to the Polyakov loop is taken complex because the Polyakov loop functional $\Phi[U]$ is complex.} $\smash{\Delta S_W^{\rho,\theta}[U]\equiv\rho e^{i\theta}\sum_{\vec{n}}\Phi[U](\vec{n})}$ which breaks the symmetry explicitly, and to study the fate of the Polyakov loop average
\beq
\langle\Phi\rangle_{\rho,\theta}\equiv \frac{\int \prod_{n,\mu}dU_\mu(n)\,\Phi[U](\vec{n})\,e^{-S_W[U]-\Delta S^{\rho,\theta}_W[U]}}{\int \prod_{n,\mu}dU_\mu(n)\,e^{-S_W[U]-\Delta S_W^{\rho,\theta}[U]}}\,,\label{eq:pert}\nonumber\\
\eeq 
in the limit $\smash{\rho\to 0^+}$.

A crucial remark is that, under the center transformation (\ref{eq:modulo}), one has
\beq
\Phi[U^g](\vec{n})=e^{-i\frac{2\pi}{3}}\Phi[U](\vec{n})\,,\label{eq:Phi_g}
\eeq
and similarly
\beq
\Phi[U^{g^\dagger}](\vec{n}) =e^{i\frac{2\pi}{3}}\Phi[U](\vec{n})\,,\label{eq:Phi_gd}
\eeq
which generalizes Eq.~(\ref{eq:gauge_inv}) for center transformations, and implies 
\beq
\Delta S_W^{\rho,\theta}[U^{g^\dagger}]=\Delta S_W^{\rho,\theta+2\pi/3}[U]\,.\label{eq:pert_g}
\eeq
One can then relabel the dummy integration variable in Eq.~(\ref{eq:pert}) as $U_\mu(n)\to U_\mu^{g^\dagger}(n)$ with a new configuration $U_\mu(n)$ periodic in all directions. Then, using that the measure and the Wilson action are invariant, whereas the Polyakov loop functional and the perturbation are transformed as in Eqs.~(\ref{eq:Phi_gd}) and (\ref{eq:pert_g}), one arrives at
\beq
\langle\Phi\rangle_{\rho,\theta}=e^{i\frac{2\pi}{3}}\langle\Phi\rangle_{\rho,\theta+2\pi/3}\,.\label{eq:idid}
\eeq
The fate of this identity as the source $\rho e^{i\theta}$ is sent to $0$, depends on whether or not the Polyakov loop admits a regular limit. For a finite system, the limit is always regular, meaning that $\lim_{\rho\to 0^+}\langle\Phi\rangle_{\rho,\theta}$ does not depend on $\theta$. Denoting this common limit as $\langle\Phi\rangle_{\rho\to 0^+}$, the identity (\ref{eq:idid}) becomes
\beq
\langle\Phi\rangle_{\rho\to 0^+}=e^{i\frac{2\pi}{3}}\langle\Phi\rangle_{\rho\to 0^+}\,,
\eeq
which implies that $\smash{\langle\Phi\rangle_{\rho\to 0^+}=0}$. In contrast, for an infinite system, the zero-source limit can be irregular in some range of temperatures. In this case, $\lim_{\rho\to 0^+}\langle\Phi\rangle_{\rho,\theta}$ still depends on $\theta$ and the identity (\ref{eq:idid}) becomes
\beq
\langle\Phi\rangle_{\rho\to 0^+,\theta}=e^{i\frac{2\pi}{3}}\langle\Phi\rangle_{\rho\to 0^+,\theta+2\pi/3}\,,
\eeq
which does not impose any constraint on the value of $\langle\Phi\rangle_{\rho\to 0^+,\theta}$ but rather a relation between the limits for two different values of $\theta$ related by the symmetry.

Of course, this second scenario of broken center symmetry never occurs in a finite system in the sense that the function $\langle\Phi\rangle_{\rho,\theta}$ is always continuously connected to $0$ as $\rho\to 0^+$. A sufficiently large system, however, can mimic a broken center symmetry behavior, if, for a certain range of temperatures, the region in $\rho$ where the Polyakov loop approaches $0$ becomes extremely small.

\subsection{Lattice implementation}\label{sec:lattice}
On the lattice, one does not need to introduce a symmetry breaking term to study the breaking of the symmetry. This is rooted in the way in which observables are determined. Let us see how this works.

Recall first that the Polyakov loop is evaluated using a Monte-Carlo ensemble ${\cal E}$ as
\beq
\langle\Phi\rangle\simeq \frac{1}{N_{\rm conf}}\sum_{U\in {\cal E}}\Phi[U](\vec{n})\equiv\langle \Phi(\vec{n})\rangle_{\cal E}\,.\label{eq:vev}
\eeq
As we have already mentioned, $\langle\Phi\rangle$ does not depend on $\vec{n}$ due to the translation invariance of the Wilson action and the use of periodic boundary conditions. In practice, this should also be satisfied within errors by the Monte-Carlo average $\langle\Phi(\vec{n})\rangle_{{\cal E}}$, but there could remain a slight dependence on $\vec{n}$. In this case, it is convenient to evaluate the Polyakov by further averaging over $\vec{n}$: 
\beq
\langle\langle\Phi\rangle\rangle_{{\cal E}}\equiv\frac{1}{L_1L_2L_3}\sum_{\vec{n}} \langle\Phi(\vec{n})\rangle_{{\cal E}}\,.
\eeq
In what follows, we stick to the single average $\langle\Phi\rangle_{{\cal E}}$ but all results discussed below apply also to the double average $\langle\langle\Phi\rangle\rangle_{{\cal E}}$.

Let us then denote by ${\cal E}^{g}$ the ensemble of configurations obtained by applying the transformation $g$ to the configurations of the original Monte-Carlo ensemble ${\cal E}$:
\beq
{\cal E}^{g}\equiv\{U_\mu^{g}\,|\,U_\mu\in {\cal E}\}\,.
\eeq
This transformed ensemble is just a convenient mathematical construct that allows one to perform a change of variables in Eq.~(\ref{eq:vev}). Indeed, it should be clear that the number of configurations in ${\cal E}^{g}$ is the same as in ${\cal E}$, and, as $U_\mu$ browses the ensemble ${\cal E}^{g}$, $U_\mu^{g^\dagger}$ browses the ensemble ${\cal E}$. One can then write
\beq
\langle \Phi\rangle_{\cal E} & = & \frac{1}{N_{\rm conf}}\sum_{U\in {\cal E}^{g}}\Phi[U^{g^\dagger}](\vec{n})\nonumber\\
& = & \frac{1}{N_{\rm conf}}\sum_{U\in {\cal E}^{g}}e^{i\frac{2\pi}{3}}\Phi[U](\vec{n})\nonumber\\
& = & \frac{e^{i\frac{2\pi}{3}}}{N_{\rm conf}}\sum_{U\in {\cal E}^{g}}\Phi[U](\vec{n})=e^{i\frac{2\pi}{3}}\langle\Phi\rangle_{{\cal E}^{g}}\,,\label{eq:steps}
\eeq
where we have used Eq.~(\ref{eq:Phi_gd}). In summary:
\beq
\langle \Phi\rangle_{\cal E}=e^{i\frac{2\pi}{3}}\langle\Phi\rangle_{{\cal E}^{g}}\,\,,\label{eq:central}
\eeq
which, to some extent, can be put into correspondance with Eq.~(\ref{eq:idid}).

From here, the discussion unfolds according to two possible scenarios:
\begin{itemize}
\item If the symmetry associated to (\ref{eq:V_transfo}) is not broken dynamically, the Monte-Carlo ensemble reflects this symmetry. That is, for most of the configurations $U_\mu(n)$ in ${\cal E}$, one can find a configuration close to $U_\mu^g(n)$ and another one close to $U_\mu^{g^\dagger}(n)$. In other words, the Monte-Carlo ensemble is approximately invariant under $g$: $\smash{{\cal E}^g\simeq{\cal E}}$. In this case, Eq.~(\ref{eq:central}) becomes
\beq
\langle \Phi\rangle_{{\cal E}}\simeq e^{i\frac{2\pi}{3}}\langle \Phi\rangle_{{\cal E}}\,.\label{eq:constraint_Phi}
\eeq
This is  a constraint on the value taken by $\langle\Phi\rangle_{{\cal E}}$ in the symmetric phase,  which eventually implies $\smash{\langle\Phi\rangle_{{\cal E}}\simeq 0}$.

\item In contrast, if the symmetry is broken spontaneously, the system explores only part of the configurations and there is no reason for ${\cal E}$ to be invariant under $g$, not even approximately: $\smash{{\cal E}^g\neq {\cal E}}$. In this case, Eq.~(\ref{eq:central}) is not a constraint on the value of $\langle\Phi\rangle_{{\cal E}}$ but rather a relation between the averaged Polyakov loops over two different patches of the configuration space, connected by the transformation (\ref{eq:V_transfo}).
\end{itemize}

The above discussion shows that the averaged Polyakov loop plays the role of an order parameter for center symmetry. One may advance, however, that this applies only to an infinite system and that, for a finite system, the second scenario, when it applies, just reflects a limitation of the sampling. Indeed, the evolution being ergodic, if one waits long enough, the Monte-Carlo ensemble should instead follow the first scenario. Still, a large enough system can mimic the breaking of ergodicity, if, in some range of temperatures, the possibility of transitioning to other regions of the configuration space becomes highly improbable. In this case, the time that would be needed to observe an ensemble complying with the first scenario, although finite, becomes extremely large, larger than any simulation time that one might conceive.

So, even though for finite systems there is strictly speaking no broken center symmetry, if the system is large enough, one may still approach a broken phase behavior, either from the study of how the system reacts to a small symmetry-breaking source, or from the nature of the dynamically generated Monte-Carlo ensemble.

It should also be noted that, in practice and within a single lattice configuration, the distribution of the Polyakov over the lattice sites differs in the symmetric and broken phases. In particular, for the symmetric phase, the phase of the Polyakov loop is equally distributed among the sectors, so the lattice average favors a vanishing Polyakov loop. On the other hand, in the broken phase, there is a preference for one of the sectors, hence providing a non-zero Polyakov loop average over the lattice, see \cite{Gattringer:2010ms} for details.

\subsection{Charge conjugation}\label{sec:charge}
Let us close this section by recalling that the Wilson action is also invariant under charge conjugation defined on the links as
\beq
U_\mu(n)\to U_\mu^C(n)\equiv U_\mu^*(n)\,.\label{eq:C0}
\eeq
Following a similar reasonning as before, we have
\beq
\Phi[U^C]=\Phi[U^*]=\Phi[U]^*\,,
\eeq
and thus that
\beq
\langle \Phi\rangle_{\cal E}=\langle \Phi \rangle^*_{{\cal E}^C}\,,\label{eq:Cc}
\eeq
where ${\cal E}^C=\{U_\mu^C\,|\,U_\mu\in {\cal E}\}$.

If charge conjugation invariance is not broken, the Monte-Carlo ensemble should be approximately invariant under $C$. In this case, we denote the ensemble by ${\cal E}_0$, such that ${\cal E}_0^C\simeq {\cal E}_0$. The identity (\ref{eq:Cc}) then becomes
\beq
\langle \Phi\rangle_{{\cal E}_0}=\langle \Phi \rangle^*_{{\cal E}_0}\,,
\eeq
which implies that $\langle\Phi[U]\rangle_{{\cal E}_0}$ is real,\footnote{The label $0$ in ${\cal E}_0$ actually refers to the fact that the phase of the averaged Polyakov loop is $0$ in this case.} in line with its physical interpretation as $e^{-\Delta F/T}$. 

Consider now two other ensembles defined by
\beq
{\cal E}_+\equiv {\cal E}_0^{g^\dagger} \quad {\rm and} \quad {\cal E}_-\equiv {\cal E}_0^{g}\,.
\eeq
In the center symmetric phase, because $\smash{{\cal E}_0^g\simeq {\cal E}_0}$, these ensembles are essentially the same as ${\cal E}_0$. In the phase of broken center symmetry, however, they are distinct from ${\cal E}_0$ but provide equally valid Monte-Carlo ensembles\footnote{By this, we mean that the configurations in ${\cal E}_\pm$ give the same value for the Wilson action than the configurations in ${\cal E}_0$.}. The ensemble ${\cal E}_+$ (resp. ${\cal E}_-$) is not invariant under $C$ but rather under the combined action of $g$, $C$ and $g^\dagger$ (resp. $g^\dagger$, $C$ and $g$) which we denote as $g^\dagger\cdot C\cdot g$ (resp. $g\cdot C\cdot g^\dagger$), where the $\cdot$ refers to the composition of transformations.\footnote{The need for this notation stems from the fact that, in the SU($3$) case, and contrary to the SU($2$) case, $C$ is not an element of SU($3$) and, therefore, the combination of $g$, $C$ and $g^\dagger$ is not just a matrix product in SU($3$).} Indeed:
\beq
{\cal E}_+^{g^\dagger\cdot C\cdot g}\equiv (({\cal E}_+^{g})^C)^{g^\dagger}=({\cal E}_0^C)^{g^\dagger}\simeq {\cal E}_0^{g^\dagger}={\cal E}_+\,.\label{eq:approx}
\eeq
Similarly, one shows that $\smash{{\cal E}_-^{g\cdot C\cdot g^\dagger}\simeq {\cal E}_-}$.

Now, we write
\beq
\Phi[U^{(g^\dagger\cdot C\cdot g)}] & = & \Phi[((U^{g})^C)^{g^\dagger}]\nonumber\\
& = & e^{i2\pi/3}\Phi[(U^{g})^C]\nonumber\\
& = & e^{i2\pi/3}\Phi[U^{g}]^*\nonumber\\
& = & e^{i4\pi/3}\Phi[U]^*\,.\label{eq:jj}
\eeq
From Eq.~(\ref{eq:jj}), one deduces
\beq
\langle \Phi\rangle_{\cal E}=e^{i4\pi/3}\langle \Phi \rangle^*_{{\cal E}^{g^\dagger\cdot C\cdot g}}\,,\label{eq:C}
\eeq
which, for the ensemble ${\cal E}_+$ obeying (\ref{eq:approx}), becomes
\beq
\langle \Phi\rangle_{{\cal E}_+}=e^{i4\pi/3}\langle \Phi \rangle^*_{{\cal E}_+}\,,\label{eq:C}
\eeq
and tells that the phase of $\langle \Phi\rangle_{{\cal E}_+}$ is fixed to $2\pi/3$ modulo $\pi$. Of course, a similar reasoning leads to the conclusion that the phase of $\langle \Phi\rangle_{{\cal E}_-}$ is fixed to $-2\pi/3$ modulo $\pi$.

Deep in the broken phase, the Monte-Carlo ensembles will always be of type ${\cal E}_0$, ${\cal E}_+$ or ${\cal E}_-$, and, by applying center transformations, we can always choose to work with an ensemble of type ${\cal E}_0$.\footnote{Whenever the system is coupled to dynamical quarks, center-symmetry is explicitly broken. In the absence of a baryonic chemical potential, charge conjugation is manifest and the configurations that are favored are those of the type ${\cal E}_0$. One way to reach configurations of the type ${\cal E}_\pm$ is to introduce an imaginary baryonic chemical potential equal to $\pm i2\pi T$.}

 \section{Link correlators as order parameters}\label{sec:links}
We have just recalled why the Polyakov loop average is an observable that plays the role of an order parameter for center symmetry. The two crucial ingredients are, first, that the Polyakov loop functional (\ref{eq:Phi_func}) transforms in a simple (linear) way under the symmetry, see Eq.~(\ref{eq:Phi_g}), and, second, that the Monte-Carlo ensemble reflects the symmetry in the symmetric phase, in the sense that ${\cal E}^g\simeq {\cal E}$, see the discussion in Sec.~\ref{sec:lattice}. In what follows, we would like to construct alternative order parameters from averages of non-gauge-invariant functionals.

\subsection{Link correlators}
The basic example of a non-gauge-invariant functional is a tensor products of links
\beq
U_{\mu_1}(n_1)\otimes\cdots\otimes U_{\mu_p}(n_p)\,,
\eeq
the simplest of which is of course the single link $U_\mu(n)$. These functionals transform linearly under the action of ${\cal G}_0$ and ${\cal G}$.

It should be stressed, however, that the corresponding averages are usually not defined using the Monte-Carlo ensemble ${\cal E}$ but, rather, using a gauge-fixed ensemble ${\cal E}_{\rm gf}$, which we shall define more precisely below:
\beq
& & \langle U_{\mu_1}(n_1)\otimes\cdots\otimes U_{\mu_p}(n_p)\rangle_{{\cal E}_{\rm gf}}\nonumber\\
& & \hspace{1.0cm}\equiv\,\frac{1}{N_{\rm conf}}\sum_{U\in {\cal E}_{\rm gf}}U_{\mu_1}(n_1)\otimes\cdots\otimes U_{\mu_p}(n_p)\,.\label{eq:corr}
\eeq
Following the same steps as for the Polyakov loop, and exploiting the fact that the links transform linearly under center transformations, one can deduce linear relations obeyed by the link correlators over a priori distinct ensembles. For instance, let us work out explicitly the case of the link average. By definition\footnote{Note that, in general, the average of SU(3) matrices is not a SU(3) matrix.}
\beq
\langle U_\mu(n)\rangle_{{\cal E}_{\rm gf}}\equiv\frac{1}{N_{\rm conf}}\sum_{U\in{\cal E}_{\rm gf}} U_\mu(n)\,,
\eeq
and, as for the Polyakov loop, we can combine the ensemble with a lattice average
\beq
\langle\langle U_\mu\rangle\rangle_{{\cal E}_{\rm gf}}\equiv \frac{1}{L_1L_2L_3L_4}\sum_{n} \langle U_\mu(n)\rangle_{{\cal E}_{\rm gf}}\,,
\eeq
We now denote by ${\cal E}_{\rm gf}^{g}$ the ensemble of configurations obtained by applying the transformation $g$ to the configurations of the original gauge-fixed ensemble ${\cal E}_{\rm gf}$:
\beq
{\cal E}_{\rm gf}^{g}=\{U_\mu^{g}\,|\, U_\mu\in {\cal E}_{\rm gf}\}\,.
\eeq
It should be clear that ${\cal E}_{\rm gf}$ and ${\cal E}_{\rm gf}^{g}$ contain the same number of configurations and that, as $U_\mu$ browses ${\cal E}_{\rm gf}^{g}$, $U_\mu^{g^\dagger}$ browses ${\cal E}_{\rm gf}$. One can then write
\beq
\langle U_\mu(n)\rangle_{{\cal E}_{\rm gf}} & = & \frac{1}{N_{\rm conf}}\sum_{U\in{\cal E}^{g}_{\rm gf}} U^{g^\dagger}_\mu(n)\nonumber\\
& = & \frac{1}{N_{\rm conf}}\sum_{U\in{\cal E}^{g}_{\rm gf}} g^\dagger(n)U_\mu(n)g(n+\hat\mu)\nonumber\\
& = & g^\dagger(n)\left[\frac{1}{N_{\rm conf}}\sum_{U\in{\cal E}^{g}_{\rm gf}} U_\mu(n)\right]g(n+\hat\mu)\,,\label{eq:steps2}\nonumber\\
\eeq
and thus
\beq
\langle U_\mu(n)\rangle_{{\cal E}_{\rm gf}}=g^\dagger(n)\,\langle U_\mu(n)\rangle_{{\cal E}^{g}_{\rm gf}}\,g(n+\hat\mu)\,.\label{eq:ididid}
\eeq
Similar relations can be deduced for higher link correlators. 

All these relations look rather similar to Eq.~(\ref{eq:central}) and then, it is tempting to deduce that the link correlators can play the role of order parameters for center symmetry. However, the possibility to use these quantities as order parameters relies on whether or not the gauge-fixed ensemble obeys ${\cal E}_{\rm gf}^{g}\simeq {\cal E}_{\rm gf}$ in the symmetric phase. 

As we now argue, for generic gauges, ${\cal E}_{\rm gf}$ and ${\cal E}_{\rm gf}^{g}$ are distinct, even in the symmetric phase, as they correspond to different gauges. In this case, Eq.~(\ref{eq:ididid}) relates the link average in two different gauges but in no way can fix the value of the link average in any of the two gauges. For specific types of gauges, however, and for specific associated center transformations $\tilde g$, the ensembles ${\cal E}_{\rm gf}$ and ${\cal E}_{\rm gf}^{{\tilde g}}$ are made of link configurations in the same gauge. Moreover, for specific choices of the Gribov copies that are potentially present in those gauges, the ensembles ${\cal E}_{\rm gf}$ and ${\cal E}_{\rm gf}^{{\tilde g}}$ are approximately equal in the symmetric phase. In this case, Eq.~(\ref{eq:ididid}), for the specific choice of center transformation $\tilde g$, becomes a constraint for the link average in that gauge, which then becomes a potential order parameter for center symmetry.

\subsection{Gauge fixing}\label{sec:gf}
Before proceeding with the proof, let us recall how the gauge fixing is implemented on the lattice. 

A given gauge is specified by the choice of a functional $F[U]$ defined over the link configurations. A gauge fixing within that gauge is obtained by replacing each configuration $\smash{U_\mu\in {\cal E}}$ of the original Monte-Carlo ensemble by a local maximum of $F[U]$ along the ${\cal G}_0$-orbit of $U_\mu$. In other words, for any fixed $\smash{U_\mu\in {\cal E}}$, one looks for a local maximum of the function $g_0\mapsto F[U^{g_0}]$ over ${\cal G}_0$.  By construction, the gauge-fixed configurations  $U_\star$ are such that
\beq
F[U_\star^{g_0}]<F[U_\star]\,,\label{eq:max}
\eeq
for any $g_0\in {\cal G}_0$ close to the identity transformation. 

It is important to keep in mind that the maximization is done along the ${\cal G}_0$-orbits and not along the ${\cal G}$-orbits. This is because, despite the fact that the transformations in ${\cal G}$ write mathematically as gauge transformations, they correspond to actual physical symmetries, and not redundancies of the description in terms of gauge fields. Nevertheless, it will be important below to study the transformation properties of $F[U]$ under the action of ${\cal G}$. This will naturally lead us to consider $F[U^g]$ for $\smash{g\in {\cal G}}$.

There are some restrictions on the functional $F[U]$. First, it should not be invariant under the action of ${\cal G}_0$, otherwise there are no local maxima (or all configurations are degenerate maxima) and gauge fixing is not possible. Second, there should be at least one maximum per ${\cal G}_0$-orbit. In fact, it is very often the case that there are actually many local maxima along each ${\cal G}_0$-orbit, known as Gribov copies. In this case, there exist various possible gauge fixings in a given gauge, depending on which copy is selected on each orbit. In what follows, we denote generically by ${\cal E}_{\rm gf}$ any such choice of copies along the orbits of the link configurations in the original Monte-Carlo ensemble ${\cal E}$.

\subsection{Generic gauge}\label{sec:gg}
We first consider a generic gauge,\footnote{By generic, we mean that we do not assume any special properties of the functional $F[U]$ other than the minimal ones listed above.} associated to a functional $F[U]$. Given a gauge-fixed ensemble ${\cal E}_{\rm gf}$ in this gauge, we would like to argue that the transformed ensemble ${\cal E}_{\rm gf}^{g}$ corresponds to a gauge-fixed ensemble in a different gauge, and thus that the two ensembles cannot coincide in this generic setting, not even in the center-symmetric phase. 

To this aim, let us show that the transformed ensemble ${\cal E}_{\rm gf}^{g}$ that enters Eq.~(\ref{eq:ididid}) is made of all the configurations that maximize the functional $F[U^{g^\dagger}]$ (seen as a functional of $U$ for a fixed $g$) along the ${\cal G}_0$-orbits of ${\cal E}^{g}$. To see this, consider a configuration $U_\star$ that maximizes $F[U]$, see Eq.~(\ref{eq:max}), and let us evaluate $F[U^{g^\dagger}]$ in the vicinity of $U_\star^g$ along its ${\cal G}_0$-orbit. This means that we have to evaluate
\beq
F[((U_\star^{g})^{g_0})^{g^\dagger}]=F[U_\star^{g^\dagger g_0g}]\,.
\eeq
Now, because $\smash{\tilde g_0\equiv g^\dagger g_0g}$ is periodic and close to the identity,\footnote{The periodicity along the spatial directions is trivial. Along the temporal direction, we write,
\begin{eqnarray*}
\tilde g_0(n+L_4\hat 4) & = & g^\dagger(n+L_4\hat 4)g_0(n+L_4\hat 4)g(n+L_4\hat 4)\\
& = & e^{-i2\pi/3}g^\dagger(n)g_0(n)g(n)e^{i2\pi/3}\nonumber\\
& = & g^\dagger(n)g_0(n)g=\tilde g_0(n)\,.
\end{eqnarray*}
} we can use Eq.~(\ref{eq:max}) to deduce that
\beq
F[((U_\star^{g})^{g_0})^{g^\dagger}]=F[U_\star^{\tilde g_0}]< F[U_\star]=F[(U_\star^{g})^{g^\dagger}]\,,
\eeq
for $g_0$ close to the identity. This shows that the configurations of ${\cal E}_{\rm gf}^{g}$ are (local) maxima of the functional $F[U^{g^\dagger}]$ along the orbits of ${\cal E}^{g}$.

But for a generic choice of gauge, the functional $F[U^{g^\dagger}]$ has no reason to be equal to the original one $F[U]$. Then, the configurations in ${\cal E}_{\rm gf}$ and in ${\cal E}_{\rm gf}^{g}$ are configurations in two distinct gauges. In this case, as already stated above, the identity (\ref{eq:ididid}) merely relates the correlators in these two gauges but in no way constraints their values in any of these gauges. For this reason, the link correlators in a generic gauge are not order parameters for center symmetry.

Note that this conclusion applies to correlators evaluated over a given ensemble. However, this does not  prevent one from constructing order parameters by comparing the correlators evaluated over different ensembles. One example of such construction in the standard Landau gauge is that of Ref. \cite{Silva:2016onh} where order parameters are constructed by comparing the propagators over distinct center sectors. We leave the comparison between the present approach and the one in Ref. \cite{Silva:2016onh} for a future work.

Note also that, even though the gauge-fixed ensemble ${\cal E}_{\rm gf}$ in a generic gauge is not invariant under any given center transformation $g$ in the symmetric phase, it still contains information about center symmetry. Indeed, given $U\in {\cal E}_{\rm gf}$, $U^g$ is in general not in ${\cal E}_{\rm gf}$, but it is possible to find a $g_0\in {\cal G}_0$ such that $(U^{g})^{g_0}=U^{g_0g}$ is back in ${\cal E}_{\rm gf}$ and thus ${\cal E}_{\rm gf}^{g_0g}\simeq {\cal E}_{\rm gf}$. One could then wonder what prevents one from repeating the same steps that lead to (\ref{eq:ididid}) using $g_0g$ instead of $g$ and, given that ${\cal E}_{\rm gf}^{g_0g}\simeq {\cal E}_{\rm gf}$, to obtain a center symmetry constraint in a generic gauge. The loophole is that, a priori, the transformation $g_0$ depends on $U$ and then so does $g_0g$. This prevents one from completing the last steps in (\ref{eq:ididid}) simply because the configuration-dependent transformation $g_0g$ cannot be pulled out of the sum over all configurations in Eq.~(\ref{eq:steps2}). This problem can be overcome by using the class of center-symmetric gauges that we now discuss.

\subsection{Center-symmetric gauges}\label{sec:csg}
Suppose that we were able to find a particular type of functionals $\tilde F[U]$ invariant under certain representatives $\smash{\tilde g\in\mathds{\cal G}}$ of the center transformations\footnote{Because the Wilson action is invariant under genuine gauge transformations $g_0$, there are infinitely many ways to define the center transformations: a center transformation $g$ is physically undistinguishable from $g_0g$. In a gauge-fixed setting, in contrast, when center symmetry is manifest, it is so only for certain representatives $g_0g$.}
\beq
\tilde F[U^{\tilde g}]=\tilde F[U]\,.\label{eq:Fc}
\eeq
We dub these functionals, and the associated gauges, as center-symmetric. We shall construct one explicit example in Sec.~\ref{sec:cslg}.

The discussion in the previous section applies here too, so the configurations in ${\cal E}_{\rm gf}^{\tilde g}$ maximize the functional $\tilde F[U^{\tilde g^\dagger}]$ along the $\mathcal{G}_0$-orbits of ${\cal E}^{\tilde g}$. But from Eq.~(\ref{eq:Fc}), it follows that the functional $\tilde F[U^{\tilde g^\dagger}]$ is nothing but the original functional $\tilde F[U]$. We deduce from this that, in a center-symmetric gauge, and contrary to the case of a generic gauge, the configurations  in ${\cal E}_{\rm gf}$ and ${\cal E}_{\rm gf}^{\tilde g}$  maximize the same functional (along a priori distinct orbits for the moment) and thus correspond to configurations in the same gauge.

Consider now the symmetric phase for which $\smash{{\cal E}^{\tilde g}\simeq {\cal E}}$. In this case, not only do the configurations in ${\cal E}_{\rm gf}$ and ${\cal E}_{\rm gf}^{\tilde g}$ maximize the same functional, but they do so along the same orbits, those of ${\cal E}^{\tilde g}\simeq{\cal E}$. If we assume for the moment that $\tilde F[U]$ is such that there are no Gribov copies, then we deduce that $\smash{{\cal E}_{\rm gf}^{\tilde g}\simeq {\cal E}_{\rm gf}}$. In this case, the identity (\ref{eq:ididid}), for the specific choice $g\to\tilde g$, becomes a constraint for the link correlator in the center-symmetric gauge:\footnote{Strictily speaking the equality is actually an approximate equality since ${\cal E}_{\rm gf}^{\tilde g}$ is only approximately equal to ${\cal E}_{\rm gf}$ due to the finite sample size.}
\beq
\langle U_\mu(n)\rangle_{{\cal E}_{\rm gf}}=\tilde g^\dagger(n)\,\langle U_\mu(n)\rangle_{{\cal E}_{\rm gf}}\,\tilde g(n+\hat\mu)\,.\label{eq:idd}
\eeq
This constraint fixes the value of the correlator (or of some of its components) in the symmetric phase which thus turns into a potential order parameter for center symmetry.\footnote{Of course, there is still the possibility that the correlator also receives stronger constraints from other, unbroken symmetries, in which case it might not be possible to use it as a probe for center symmetry.}

Note also that in the broken phase, we will have ${\cal E}_{\rm gf}^{\tilde g}$ distinct from ${\cal E}_{\rm gf}$, but, as we already pointed out above, contrary to what happens in the generic case, here these two ensembles contain configurations in the same gauge. These ensembles can be interpreted as distinct center symmetry broken states connected by the symmetry.

\subsection{Gribov copies}\label{sec:Gribov}

Let us now see how the discussion extends in the presence of Gribov copies. Of course, it could be that the quantity that we want to put forward as an order parameter is not much affected by the choice of copies, in which case the discussion in the previous section applies and the gauge-fixed ensemble can be assumed to be approximately $\tilde g$-invariant in the symmetric phase.

In the case where the choice of copies has a non-negligible impact on the quantity of interest, the discussion is slightly more subtle. Indeed, the fact that the ensembles ${\cal E}_{\rm gf}$ and ${\cal E}_{\rm gf}^{\tilde g}$ maximize the same functional along the same orbits is not enough to conclude that $\smash{{\cal E}_{\rm gf}^{\tilde g}\simeq {\cal E}_{\rm gf}}$. This  very property, and thus the possibility of deriving (\ref{eq:idd}) and of using the link average as an order parameter, depends crucially on the way the copies are chosen. Stated differently, in this case, having a center-symmetric gauge does not imply necessarily that the gauge fixing is center-symmetric.

Consider first the ideal situation of absolute gauge fixing where the functional $\tilde F[U]$ is assumed to have only one absolute maximum per gauge orbit and that this is the maximum selected to form the gauge-fixed ensemble. Since the configurations in ${\cal E}_{\rm gf}$ and ${\cal E}_{\rm gf}^{\tilde g}$ absolutely maximize the functional $\tilde F[U]$ along the same orbits, those of ${\cal E}^{\tilde g}\simeq{\cal E}$ in the symmetric phase, and because the absolute maximum is assumed to be unique along each ${\cal G}_0$-orbit, we deduce that  $\smash{{\cal E}_{\rm gf}^{\tilde g}\simeq {\cal E}_{\rm gf}}$ in the symmetric phase. So maximal gauge fixing in a center-symmetric gauge is itself center-symmetric.

Unfortunately, such type of gauge fixing is difficult, if not impossible to implement in practice and the lattice approach resorts instead to choosing one local maximum of the gauge-fixed functional per ${\cal G}_0$-orbit (this is sometimes referred to as ``minimal'' gauge fixing). The search for maxima is typically done using algorithms based on the steepest ascent method. In App.~\ref{app:ascent}, we show that, in the case of a center-symmetric gauge $\tilde F[U]$, the most primitive version of such algorithms is symmetry preserving: applying it to two configurations $U$ and $U^{\tilde g}$ related by a transformation $\tilde g$ that leaves $\tilde F[U]$ invariant, it gives maxima on the ${\cal G}_0$-orbits of $U$ and $U^{\tilde g}$ that are themselves related by $\tilde g$. This ensures that ${\cal E}_{\rm gf}^{\tilde g}\simeq {\cal E}_{\rm gf}$.

In practice, however,  this basic steepest ascent algorithm is upgraded using various acceleration routines which may destroy this symmetry preserving property, at least partially. It is of interest to investigate the possibility of constructing explicit symmetry preserving algorithms. However, we note that this may not be really necessary for practical purposes. Here is the argument.

 Indeed, as we have already stressed, the Monte-Carlo ensemble is generated using the Wilson action which is gauge-invariant, and, because gauge invariance cannot break dynamically, the configurations of the Monte-Carlo ensemble can be grouped in subsets of configurations that are almost along the same orbits. Then, even though ${\cal E}_{\rm gf}$ is obtained by selecting one copy per orbit, the configurations forming this ensemble can be grouped in subsets of configurations that are almost Gribov copies of each other. This clearly increases the possibility that, in the symmetric phase, some fraction of the selected maxima are approximately pairwise related by the transformation $\tilde g$ defined in Eq.~(\ref{eq:Fc}). Note that this is true irrespectively of whether we construct the gauge-fixed ensemble in the center-symmetric gauge starting from a genuine Monte-Carlo ensemble or from any other gauge-fixed ensemble in a generic gauge. This is because, as we have already mentioned, gauge-fixed ensembles in generic gauges still contain the information about center symmetry although this information might be hidden behind a configuration-dependent center transformation.

Note finally that this second argument applies strictly speaking in the scenario where the Monte-Carlo ensemble is sufficiently large. In practice, the ensemble is of a moderate size and the above expectation applies only with some degree of approximation. At the end of the day, the best that can be done regarding the question of whether or not $\smash{{\cal E}_{\rm gf}^{\tilde g}\simeq {\cal E}_{\rm gf}}$ in the symmetric phase, is to verify numerically that, for a given choice of copies, the consequences of $\smash{{\cal E}_{\rm gf}^{\tilde g}\simeq {\cal E}_{\rm gf}}$, such as Eq.~(\ref{eq:idd}), are realized, and also to try to quantify how much ${\cal E}_{\rm gf}^{\tilde g}$ resembles ${\cal E}_{\rm gf}$. This, we shall do below.

\section{The center-symmetric Landau gauge}\label{sec:cslg}

After these general considerations, let us now construct an explicit example. We define the center-symmetric Landau gauge by the functional
\beq
\tilde F[U]\equiv \sum_{n,\mu} {\rm Re}\, {\rm tr}\, g^\dagger_c(\hat\mu)\,U_\mu(n)\,,\label{eq:csLandau}
\eeq
with (recall that $L_4$ denotes the temporal extent of the lattice)
\beq
g_c(\hat\mu)=e^{-i\frac{4\pi}{3}\frac{\hat\mu_4}{L_4}\frac{\lambda^3}{2}}\,.\label{eq:13}
\eeq
We stress again that the gauge fixing associated to $\tilde F[U]$ amounts to looking for local maxima of the function $g_0\mapsto \tilde F[U^{g_0}]$ over ${\cal G}_0$.

The form of the functional (\ref{eq:csLandau}) as well as the transformation (\ref{eq:g}) below can be guessed from the continuum implementation of background Landau gauges in general and of the center-symmetric Landau gauge in particular, as we explain in App.~\ref{app:guess}.

\subsection{Center invariance}
We now claim that the functional (\ref{eq:csLandau}) is invariant under the particular center transformation:\footnote{For any other choice of center transformation, it is always possible to adapt the choice of $g_c(\hat\mu)$ in Eq.~(\ref{eq:csLandau}) to make sure that the functional $\tilde F[U]$ is invariant under that particular center transformation.}
\beq
\tilde g(n)=e^{i\pi\frac{\lambda_4}{2}}e^{i\pi\frac{\lambda_1}{2}}e^{-i\frac{n_4}{L_4}\pi\left(\lambda_3+\frac{\lambda_8}{\sqrt{3}}\right)}\,.\label{eq:g}
\eeq
Let us first check that it is indeed a center transformation. We have
\beq
\tilde g(n+L_4\hat{4}) & = & e^{i\pi\frac{\lambda_4}{2}}e^{i\pi\frac{\lambda_1}{2}}e^{-i\frac{n_4}{L_4}\pi\left(\lambda_3+\frac{\lambda_8}{\sqrt{3}}\right)}e^{-i\pi\left(\lambda_3+\frac{\lambda_8}{\sqrt{3}}\right)}\nonumber\\
& = & \tilde g(n)\exp\left\{-i\pi\left(
\begin{array}{ccc}
1+\frac{1}{3} & 0 & 0\\
0 & -1+\frac{1}{3} & 0\\
0 & 0 & -\frac{2}{3} 
\end{array}
\right)\right\}\nonumber\\
& = & e^{i\frac{2\pi}{3}}\tilde g(n)\,.
\eeq
To show that it leaves the functional $\tilde F[U]$ invariant, the trick is to write
\beq
\tilde F[U^{\tilde g}] & = & \sum_{n,\mu} {\rm Re}\, {\rm tr}\, g^\dagger_c(\mu) \tilde g(n)U_\mu(n)\tilde g^\dagger(n+\hat\mu)\nonumber\\\nonumber\\
& = &  \sum_{n,\mu} {\rm Re}\, {\rm tr}\, \tilde g^\dagger(n+\hat\mu) g^\dagger_c(\mu)\tilde g(n)U_\mu(n)\,,
\eeq
so it all boils down to showing that
\beq
\tilde g^\dagger(n+\hat\mu) g^\dagger_c(\hat\mu) \tilde g(n)=g^\dagger_c(\hat\mu)\,.\label{eq:17}
\eeq
For $\smash{\mu\neq4}$, this is trivial because $\smash{g_c(\hat\mu)=\mathds{1}}$ and $\smash{\tilde g(n+\hat\mu)=\tilde g(n)}$. For $\smash{\mu=4}$, one has to work a bit more but it can be done as we show in App.~\ref{app:check}. We will later see how this can be understood using Weyl transformations. Then, as announced,
\beq
\tilde F[U^{\tilde g}]=\tilde F[U]\,.\label{eq:sym}
\eeq
This identity is valid for any $U$, in particular for $U^{\tilde g^\dagger}$ and thus we deduce that $\tilde F[U]$ is also invariant under the center transformation $\tilde g^\dagger$, whose associated center element is $e^{-i2\pi/3}$. Alternatively, this can be deduced from the fact that Eq.~(\ref{eq:17}) also rewrites
\beq
g^\dagger_c(\hat\mu)=\tilde g(n+\hat\mu) g^\dagger_c(\hat\mu)\tilde g^\dagger(n)\,,\label{eq:28}
\eeq
an identity that we shall be using soon.

\subsection{Other symmetries}\label{sec:other}
The center-symmetric Landau gauge  functional benefits from additional invariance properties that constrain the structure of the link correlators. First, as it can trivially be checked, the functional $\tilde F[U]$ is invariant under global color rotations of the form
\beq
U_\mu(n)\to  e^{i\theta^jt^j}U_\mu(n)e^{-i\theta^jt^j}\equiv U^\theta_\mu(n)\,,\label{eq:color}
\eeq
where the $t^j$ denote the commuting generators of the algebra, here $\lambda_3/2$ and $\lambda_8/2$.
Second, it is invariant (and so is the Wilson action) under the transformation:
\beq
U_\mu(n)\to  e^{i\pi\frac{\lambda_1}{2}} U_\mu^*(n)e^{-i\pi\frac{\lambda_1}{2}} \equiv U_\mu^{\hat C}(n)\,.\label{eq:Ct}
\eeq
which combines charge conjugation as defined in Eq.~(\ref{eq:C0}) and a particular color rotation (known as Weyl transformation).\footnote{As any other physical symmetry, the action of charge conjugation on gauge-dependent objects such as links, is defined modulo a gauge transformation, so $U_\mu(n)\to U_\mu^*(n)$ or $U_\mu(n)\to g_0(n)U_\mu^*(n)g_0^\dagger(n+\hat\mu)$ are equally good definitions of charge conjugation, in the sense that they act equally on observables. In a gauge-fixed setting, it might be useful to consider one particular realization, in particular if it leaves the gauge-fixing functional invariant. This is what we have done here.} Indeed
\beq
\tilde F[U^{\hat C}] & = & \sum_{n,\mu} {\rm Re}\, {\rm tr}\, g^\dagger_c(\hat\mu) e^{i\pi\frac{\lambda_1}{2}} U_\mu^*(n)e^{-i\pi\frac{\lambda_1}{2}}\nonumber\\
& = & \sum_{n,\mu} {\rm Re}\, {\rm tr}\, g^{\rm t}_c(\hat\mu) e^{-i\pi\frac{\lambda_1}{2}} U_\mu(n)e^{i\pi\frac{\lambda_1}{2}}\nonumber\\
& = & \sum_{n,\mu} {\rm Re}\, {\rm tr}\, e^{i\pi\frac{\lambda_1}{2}}g_c(\hat\mu) e^{-i\pi\frac{\lambda_1}{2}} U_\mu(n)\,,
\eeq
so it all boils down to showing that 
\beq
e^{i\pi\frac{\lambda_1}{2}}g_c(\mu) e^{-i\pi\frac{\lambda_1}{2}}=g^\dagger_c(\mu)\,.\label{eq:54}
\eeq 
For $\smash{\mu\neq 4}$, this is obvious. For $\smash{\mu=4}$, see App.~\ref{app:check}.

Let us stress that the Monte-Carlo ensembles of type ${\cal E}_0$, see the discussion in Sec.~\ref{sec:charge}, which are approximately invariant under $C$ defined in Eq.~(\ref{eq:C0}), are also approximately invariant under $\hat C$ defined in Eq.~(\ref{eq:Ct}). This is because, a Monte-Carlo ensemble should always be approximately invariant under genuine gauge transformation, and thus, in particular, under global color rotations. Since the connection between $C$ and $\hat C$ is precisely a global color rotation, we deduce that ${\cal E}_0$ should also be invariant under $\hat C$.

Now, using that the center-symmetric Landau gauge functional $\tilde F[U]$ is invariant under $\hat C$, and using a similar reasonning as the one done above for the invariance under $\tilde g$, we deduce that for appropriately chosen copies, the gauge-fixed ensemble ${\cal E}_{0{\rm gf}}$ is invariant under $\hat C$. This remark is important because it implies that, as long as we do our gauge-fixing starting from an ensemble of type ${\cal E}_0$, we will be able to use the constraints associated to the symmetry $\hat C$, see below.

Now, suppose we wanted to do the gauge-fixing starting from the Monte-Carlo ensembles ${\cal E}_+={\cal E}_0^{\tilde g^\dagger}$ or ${\cal E}_-={\cal E}_0^{\tilde g}$, see Sec.\ref{sec:charge}. The first good news is that it is very simple to do so in the center-symmetric Landau gauge, because one just needs to consider as gauge-fixed ensembles $\smash{{\cal E}_{+{\rm gf}}\equiv{\cal E}_{0{\rm gf}}^{\tilde g^\dagger}}$ and $\smash{{\cal E}_{-{\rm gf}}\equiv{\cal E}_{0{\rm gf}}^{\tilde g}}$. Indeed, unlike what would happen in a generic gauge, these ensembles contain configurations in the same center-symmetric gauge as ${\cal E}_{0{\rm gf}}$. Moreover, the configurations in ${\cal E}_{\pm{\rm gf}}$ lie along the ${\cal G}_0$-orbits of the configurations in ${\cal E}_\pm$. These gauge-fixed ensembles are not approximately invariant under $\hat C$ but following similar steps as those in Sec.~\ref{sec:charge}, it is easily argued that they are approximately invariant under $\tilde g^\dagger\cdot\hat C\cdot \tilde g$ and $\tilde g\cdot\hat C\cdot \tilde g^\dagger$ respectively.

\subsection{Link average}
We are now fully equipped to derive center-symmetry constraints on the link average computed in the center-symmetric Landau gauge. In what follows, we slightly simplify the notation by dropping the reference to the gauge-fixing ensemble in the definition of the average:
\beq
\langle U_\mu(n)\rangle_{{\cal E}_{\rm gf}}\to \langle U_\mu(n)\rangle\,.
\eeq

According to the general discussion, if center symmetry is not broken by the gauge-fixed ensemble, we should have
\beq
\langle U_\mu(n)\rangle=\tilde g(n)\,\langle U_\mu(n)\rangle\,\tilde g^\dagger(n+\hat\mu)\,.\label{eq:24}
\eeq
This is a constraint on the possible values of the local link average in the symmetric phase, just as Eq.~(\ref{eq:constraint_Phi}) is a constraint for the value of the Polyakov loop in the symmetric phase. 

The question is now: what are the possible values for $\langle U_\mu(n)\rangle$ in the symmetric phase?  We know already one object that obeys Eq.~(\ref{eq:24}). Indeed, Eq.~(\ref{eq:28}) rewrites as
\beq
g_c(\hat\mu)= \tilde g(n)g_c(\hat\mu)\tilde g^\dagger(n+\hat\mu)\,,\label{eq:26}
\eeq
and so $g_c(\hat\mu)$ obeys the constraint (\ref{eq:24}), and more generally $\eta g_c(\hat\mu)$ with $\eta\in\mathds{C}$. Let us now show that this is the only possible solution of the symmetry contraint.

Let us introduce some useful notions. First of all, it will be convenient to decompose the links, and therefore the link correlators, along appropriately chosen color bases. The two useful choices are

\beq
U_\mu(n) & = & \sum_{\rho'\rho}U^{\rho'\rho}_\mu(n)|\rho'\rangle\langle\rho|\nonumber\\
& = & U^1_\mu(n)\mathds{1}+\sum_\kappa U^\kappa_\mu(n) t^\kappa\,,\label{eq:bases}
\eeq
where the $\rho$'s are the defining weights of the color algebra, such that\footnote{If we take for the generators $t^j$ some diagonal matrices, the $|\rho\rangle$ are nothing but the canonical vectors, with only one non-zero entry, equal to $1$.} $\smash{t^j|\rho\rangle=\rho^j|\rho\rangle}$, while the $\kappa$'s are the weights of the adjoint representation, such that $\smash{[t^j,t^\kappa]=\kappa^j t^\kappa}$. Recall that some of the adjoint weights can vanish, in which case they are denoted as $\smash{\kappa=0^{(j)}}$ associated to the diagonal generators $\smash{t^\kappa=t^j}$. The non-zero adjoint weights are called roots and denoted $\alpha$. The relation between the two bases is easily found. In particular\footnote{The relation between $U_\mu^\alpha$ and $U_\mu^{\rho\rho'}$ can also be worked out but we shall not be needing it here.}
\beq
U_\mu^1(n) & = & \frac{1}{N}\sum_\rho U^{\rho\rho}_\mu(n)\,,\\\nonumber\\
U_\mu^j(n) & = & 2\sum_\rho \rho^j\,U^{\rho\rho}_\mu(n)\,.\label{eq:change}
\eeq

\noindent{Second, it is useful to re-derive Eq.~(\ref{eq:26}) using the notion of Weyl transformations, see for instance \cite{vanEgmond:2023lfu}. We can first rewrite the transformation (\ref{eq:g}) as}
\beq
\tilde g(n)=W_{\alpha_{13}}W_{\alpha_{12}}e^{-i\frac{n_4}{L_4}4\pi\rho_1\cdot t}\,,\label{eq:g2}
\eeq
where $\smash{X\cdot t}$ is a short-hand notation for $X^j t^j$, while $\smash{\alpha_{12}=\rho_1-\rho_2}$ and $\smash{\alpha_{13}=\rho_1-\rho_3}$ denote two of the SU(3) roots expressed in terms of the SU(3) defining weights $\smash{\rho_1=(1,1/\sqrt{3})/2}$, $\smash{\rho_2=(-1,1/\sqrt{3})/2}$ and $\smash{\rho_3=(0,-1/\sqrt{3})}$. The transformations $W_{\alpha_{12}}$ and $W_{\alpha_{13}}$ are Weyl transformations whose explicit expressions (given in Ref.~\cite{vanEgmond:2023lfu}) are not very important. What we need to know is their action on the algebra. In particular
\beq
W_\alpha (X\cdot t) W^\dagger_\alpha=(R_\alpha X)\cdot t\,,
\eeq
where $R_\alpha$ is the reflection w.r.t. an axis orthogonal to the root $\alpha$. 

Let us now use these notions to re-derive the identity (\ref{eq:26}). We can write
\begin{widetext}
\beq
\tilde g(n)g_c(\hat\mu)\tilde g^\dagger(n+\hat\mu) & = & W_{\alpha_{13}}W_{\alpha_{12}}e^{-i\frac{n_4}{L_4}4\pi\rho_1^jt^j}e^{-i\frac{4\pi}{3}\frac{\hat\mu_4}{L_4}\frac{\lambda_3}{2}}e^{i\frac{n_4+\hat\mu_4}{L_4}4\pi\rho_1^jt^j}W^\dagger_{\alpha_{12}}W^\dagger_{\alpha_{13}}\nonumber\\
& = & W_{\alpha_{13}}W_{\alpha_{12}}e^{-i\frac{\hat\mu_4}{L_4}(r_c-4\pi\rho_1)\cdot t}W^\dagger_{\alpha_{12}}W^\dagger_{\alpha_{13}}\,,\nonumber\\
& = & e^{-i\frac{\hat\mu_4}{L_4}(R_{\alpha_{13}}R_{\alpha_{12}}(r_c-4\pi\rho_1))\cdot t}\,,
\eeq
\end{widetext}
where $\smash{r_c=(4\pi/3,0)}$. Now, it is easily checked that $R_{\alpha_{13}}R_{\alpha_{12}}$ is nothing but the rotation ${\cal R}$ by an angle $2\pi/3$ and that ${\cal R}(r_c-4\pi\rho_1)=r_c$. Thus
\beq
\tilde g(n)g_c(\hat\mu)\tilde g^\dagger(n+\hat\mu) =e^{-i\frac{\hat\mu_4}{L_4}r_c\cdot t}=g_c(\hat\mu)\,,
\eeq
as announced.

Let us now go back to our problem of identifying the most general form of $\langle U_\mu(n)\rangle$ compatible with the constraint (\ref{eq:24}). We note that color invariance (\ref{eq:color}) imposes $\smash{(e^{i\theta\cdot(\rho-\rho')}-1)\langle U_\mu^{\rho'\rho}(n)\rangle=0}$ and thus, without loss of generality, we can write
\beq
\langle U_\mu(n)\rangle=\sum_\rho \langle U^{\rho\rho}_\mu(n)\rangle|\rho\rangle\langle\rho|\,.\label{eq:diagonal}
\eeq
We then note that
\beq
\tilde g(n)|\rho\rangle & = & W_{\alpha_{13}}W_{\alpha_{12}}e^{-i\frac{n_4}{L_4}4\pi\rho_1\cdot t}|\rho\rangle\nonumber\\
& = & e^{-i\frac{n_4}{L_4}4\pi\rho_1\cdot \rho}W_{\alpha_{13}}W_{\alpha_{12}}|\rho\rangle\nonumber\\
& = & e^{-i\frac{n_4}{L_4}4\pi\rho_1\cdot \rho}|R_{\alpha_{13}}R_{\alpha_{12}}\rho\rangle\nonumber\\
& = & e^{-i\frac{n_4}{L_4}4\pi\rho_1\cdot \rho}|{\cal R}\rho\rangle\,,
\eeq
where we used that
\beq
W^\dagger_\alpha |\rho\rangle=|R_\alpha\rho\rangle\,.
\eeq
It follows that
\beq
& & \tilde g(n)\langle U_\mu(n)\rangle \tilde g^\dagger(n+\hat\mu)\nonumber\\
& & \hspace{0.5cm}=\,\sum_\rho e^{i\frac{\hat\mu_4}{L_4}4\pi\rho_1\cdot \rho}\langle U^{\rho\rho}_\mu(n)\rangle|{\cal R}\rho\rangle\langle{\cal R}\rho|\nonumber\\
& & \hspace{0.5cm}=\,\sum_\rho e^{i\frac{\hat\mu_4}{L_4}4\pi\rho_1\cdot {\cal R}^{-1}\cdot\rho}\langle U^{{\cal R}^{-1}\cdot\rho{\cal R}^{-1}\cdot\rho}_\mu(n)\rangle|\rho\rangle\langle\rho|\,.\nonumber\\
\eeq
In other words, the center transformation acts on the diagonal components of the link average as
\beq
\langle U^{\rho\rho}_\mu(n)\rangle\to e^{i\frac{\hat\mu_4}{L_4}4\pi\rho_1\cdot {\cal R}^{-1}\cdot\rho}\langle U^{{\cal R}^{-1}\cdot\rho{\cal R}^{-1}\cdot\rho}_\mu(n)\rangle\,.
\eeq
Making the different $\rho$'s explicit, this becomes
\beq
\langle U^{\rho_1\rho_1}_\mu(n)\rangle & \to & e^{-i\frac{\hat\mu_4}{L_4}\frac{2\pi}{3}}\langle U^{\rho_3\rho_3}_\mu(n)\rangle\,,\label{eq:69}\\
\langle U^{\rho_2\rho_2}_\mu(n)\rangle & \to & e^{i\frac{\hat\mu_4}{L_4}\frac{4\pi}{3}}\langle U^{\rho_1\rho_1}_\mu(n)\rangle\,,\\
\langle U^{\rho_3\rho_3}_\mu(n)\rangle & \to & e^{-i\frac{\hat\mu_4}{L_4}\frac{2\pi}{3}}\langle U^{\rho_2\rho_2}_\mu(n)\rangle\,.\label{eq:71}
\eeq
So, the constraint (\ref{eq:24}) becomes
\beq
e^{-i\frac{\hat\mu_4}{L_4}\frac{2\pi}{3}}\langle U^{\rho_3\rho_3}_\mu(n)\rangle & = & \langle U^{\rho_1\rho_1}_\mu(n)\rangle\,,\\
e^{i\frac{\hat\mu_4}{L_4}\frac{4\pi}{3}}\langle U^{\rho_1\rho_1}_\mu(n)\rangle & = & \langle U^{\rho_2\rho_2}_\mu(n)\rangle\,,\\
e^{-i\frac{\hat\mu_4}{L_4}\frac{2\pi}{3}}\langle U^{\rho_2\rho_2}_\mu(n)\rangle & = & \langle U^{\rho_3\rho_3}_\mu(n)\rangle\,,
\eeq
from which one deduces
\beq
\langle U_\mu(n)\rangle & = & \langle U^{\rho_3\rho_3}_\mu(n)\rangle\nonumber\\
& \times & \Big[e^{-i\frac{\hat\mu_4}{L_4}\frac{2\pi}{3}}|\rho_1\rangle\langle\rho_1|+e^{i\frac{\hat\mu_4}{L_4}\frac{2\pi}{3}}|\rho_2\rangle\langle\rho_2|+|\rho_3\rangle\langle\rho_3|\Big].\nonumber\\
\eeq
It is easily checked that the expression between brackets is nothing but $g_c(\mu)$ as given in Eq.~(\ref{eq:13}), and so
\beq
\langle U_\mu(n)\rangle=\eta\,g_c(\hat\mu)=\eta\,e^{-i\frac{4\pi}{3}\frac{\mu_4}{L_4} \frac{\lambda^3}{2}}\,,\label{eq:eta_gc}
\eeq
with $\eta\in\mathds{C}$. We have thus shown that, up to a numerical pre-factor $\eta$,\footnote{Because the elements of a unitary matrix have moduli bounded by $1$, so is the case of the elements of the average. From this we deduce that $|\eta|\leq 1$. We will show below that $\smash{\eta\in\mathds{R}}$.} the matrix structure of  $\langle U_\mu(n)\rangle$ is nothing but $g_c(\hat\mu)$.

In particular, for a temporal link average in the symmetric phase we find\footnote{As before, the fact that the link average does not depend on $n$ is a consequence of translation invariance, which holds true owing to the translation invariance of the Wilson action and the use of periodic boundary conditions. Within a given Monte-Carlo evaluation of the link average there will remain a spurious dependence on $n$. In this case, it might be convenient to define the link average by further averaging over the lattice sites $$\langle\langle U_4\rangle\rangle_{{\cal E}}\equiv\frac{1}{L_1L_2L_3L_4}\sum_{n} \langle U_4(n)\rangle_{{\cal E}}\,.$$}
\beq
\langle U_4(n)\rangle =
\eta \left( 
\begin{array}{ccc}
e^{- i\frac{2\pi}{3 \, L_4}} & 0 & 0 \\
0 & e^{i\frac{2\pi}{3 \, L_4}} & 0 \\
0 & 0 & 1
\end{array}
\right).\label{eq:1pt}
\eeq
We will see below that $\smash{\eta\in\mathds{R}}$ and, because $\smash{{\rm det}\,\langle U_4(n)\rangle=\eta^3}$, we can rewrite Eq.~(\ref{eq:1pt}) as
\beq
\frac{\langle U_4(n)\rangle}{({\rm det}\,\langle U_4(n)\rangle)^{1/3}}=\left( 
\begin{array}{ccc}
e^{- i\frac{2\pi}{3 \, L_4}} & 0 & 0 \\
0 & e^{i\frac{2\pi}{3 \, L_4}} & 0 \\
0 & 0 & 1
\end{array}
\right).\label{eq:38}
\eeq 
The LHS of this equation is a quantity that can be easily evaluated on the lattice (and does not require renormalization). Any deviation from the value on the RHS signals the dynamical breaking of center-symmetry. We have thus shown that, within the particular gauge considered here, a local, gauge-dependent quantity, such as the link average, can be used as an order parameter for center symmetry. 

We stress once more that the color diagonal structure (\ref{eq:diagonal}) of $\langle U_\mu(n)\rangle$ is in fact a consequence of the invariance under global color rotations (\ref{eq:color}) and applies to both the confined and deconfined phases. Similarly, over a $\hat C$-invariant ensemble, see Sec.~\ref{sec:other}, we have
\beq
W_{\alpha_{12}}\langle U_4(n)\rangle^*W^\dagger_{\alpha_{12}}=\langle U_4(n)\rangle\,,
\eeq
\vglue1mm
\noindent{in both the confined and deconfined phases. This implies} 
\beq
\langle U_4^{\rho_1\rho_1}\rangle=\langle U_4^{\rho_2\rho_2}\rangle^*\,,\label{eq:c12}
\eeq 
and
\beq
\langle U_4^{\rho_3\rho_3}\rangle\in\mathds{R}\,.\label{eq:c3}
\eeq 
As a consequence, the determinant of $\langle U_4\rangle$ should always be real.

Similarly, for a $\smash{g\cdot \hat C\cdot g^\dagger}$-invariant ensemble, see above, it can be shown that the charge conjugation constraints are
\beq
\langle U_4^{\rho_1\rho_1}\rangle=e^{-i\frac{4\pi}{3L_4}}\langle U_4^{\rho_1\rho_1}\rangle^*\,,
\eeq 
and
\beq
\langle U_4^{\rho_2\rho_2}\rangle=e^{i\frac{2\pi}{3L_4}}\langle U_4^{\rho_3\rho_3}\rangle^*\,.
\eeq 
It follows that the first diagonal element lies in the direction of $e^{-i\frac{2\pi}{3L_4}}$, while the other two have the same modulus and their product has a phase fixed to $e^{i\frac{2\pi}{3L_4}}$. The product of the phases of the three diagonal elements is then real, ensuring once again that the determinant of $\langle U_4\rangle$ is real. Similar conclusions apply for a $\smash{g\cdot \hat C\cdot g^\dagger}$-invariant ensemble, but this time the second diagonal element lies in the direction of $e^{i\frac{2\pi}{3L_4}}$, while the product of the other two has a phase fixed to $e^{-i\frac{2\pi}{3L_4}}$.

What center symmetry does is to further specify the relation between the various diagonal coefficients $\smash{\langle U_4^{\rho\rho}\rangle}$ in the symmetric phase: up to a factor $\eta$, the diagonal elements are  $e^{-i/L_4 2\pi/3}$, $e^{i/L_4 2\pi/3}$ and $1$, and this for the three types of ensembles. Notice also that $\eta$ being equal to $U_4^{\rho_3\rho_3}$, it needs to be real from Eq.~(\ref{eq:c3}), as announced above.

\subsection{Results \label{SecResults}}

Let us confront these theoretical expectations with lattice simulations for pure Yang-Mills theory.

The simulations reported here use the Wilson gauge action at $\smash{\beta=6.0}$, with gauge fixing relying on the functional (\ref{eq:csLandau}) and using a Fourier accelerated steepest descent method as optimizing algorithm, with the help of Chroma \cite{Edwards} and PFFT \cite{Pippig} libraries.  For this $\beta$ value the simulations at zero temperature  \cite{Oliveira:2012eh} suggest that finite lattice effects are small. The simulations consider two temperatures, one above  $T_c$ and another one below $T_c$ for gauge fixed ensembles with 100 gauge configurations. The lattice sizes used are $\smash{64^3 \times 6}$ and $\smash{64^3 \times 8}$. Measuring the temperature as $\smash{T = 1 / a L_4}$,  with $\smash{1/a = 1.943}$ GeV, the temperature for the $\smash{64^3 \times 6}$ lattice  is $\smash{T = 324}$ MeV, while the  $\smash{64^3 \times 8}$ lattice corresponds to a $\smash{T = 243}$ MeV. We recall that, in these units, for pure Yang-Mills theory, the temperature at which deconfinement occurs is $\sim 270$ MeV. The starting configurations for the gauge fixing algorithm are taken to be those in the Landau gauge which were already available from previous projects. Once a gauge-fixed ensemble is generated, we can generate two other ensembles in the same gauge by applying the transformations $\tilde g$ and $\tilde g^\dagger$. As explained above, in the low temperature phase, we expect these three ensembles to be essentially the same, see in particular Fig.~\ref{fig:sectors} below, while in the high temperature phase, the three ensembles correspond to three different center sectors, characterized by different transformation properties under charge conjugation, see Sect.~\ref{sec:other}.

In order to illustrate the outcome of the simulations, we consider both the ensemble average link at a single lattice size, denoted as $\langle \,\dots\, \rangle$, and its mean value over the lattice and gauge ensemble, denoted as $\langle\langle \,\dots\, \rangle\rangle$. In what concerns the statistical errors, they were evaluated either by a single jackknife elimination or the bootstrap method with a 68\% confidence level. When compared both methods produce essentially the same estimations for the errors. The results for the average over lattice and gauge ensemble have smaller statistical errors. A full numerical account of these averages, including the statistical errors, is reported in App.~\ref{app:averages}. Here we just show a simplified version of the results.

\begin{center}
\begin{figure*}[t]
\includegraphics[height=0.23\textheight]{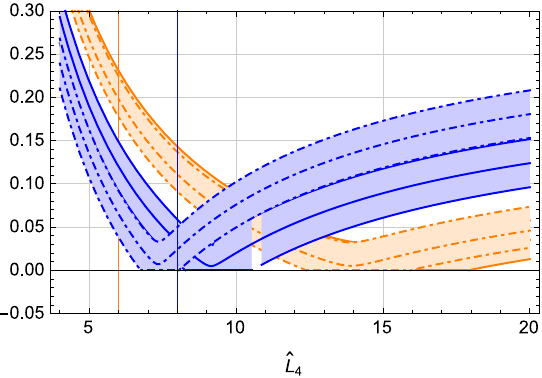}\qquad\includegraphics[height=0.23\textheight]{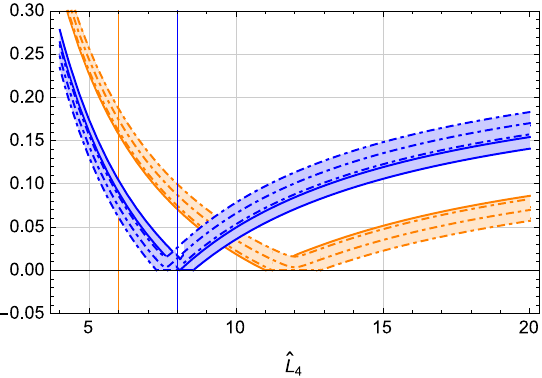}
\caption{Left plot: The distances $|M_{11}-\exp\{-i/\hat{L}_4 2\pi/3\}|$ and $|M_{22}-\exp\{i/\hat{L}_4 2\pi/3\}|$ at site $n=0$ (plain and dash-dotted curves), as a function of an auxiliary parameter $\hat{L}_4$ that takes the lattice value $L_4$ at the vertical line. Low temperature data in blue ($\smash{L_4=8}$) and high temperature data in orange ($\smash{L_4=6}$). Right plot: the same for the average of $\langle U_4(n)\rangle/({\rm det}\,\langle U_4(n)\rangle)^{1/3}$ over ten sites.}
\label{fig:dist}
\end{figure*}
\end{center}

\vglue-11mm

Let us start by the ensemble averages of the temporal link at a given lattice site. We shall take the origin of the lattice as an example but we have checked that the observations to be made below apply to the other sites as well. For the temperature above the transition that uses the $64^3 \times 6$ lattice it
follows that, for each type of ensemble,
\begin{widetext}

\begin{eqnarray}
& & 
\langle U_4 \rangle \approx
0.881
~ \left(
\begin{array}{lll}
0.996 ~ e^{-0.140} &      0.030~ e^{i 0.556} &   0.059 ~ e^{-i 1.187} \\
   0.018 ~ e^{i 1.648} &    0.995 ~ e^{i 0.150} &  0.004 ~ e^{ i 1.833} \\
   0.055 ~ e^{ -i 2.170 } &    0.007 ~ e^{i 1.894}&  1.005 ~ e^{- i 0.010 }
\end{array}
\right),
\\
& &
\langle U^{\tilde g}_4 \rangle \approx
0.881
~ \left(
\begin{array}{ccc}
1.005 ~ e^{- i 0.360} &   0.055 ~  e^{-i 1.472} &  0.007 ~ e^{-i 0.026}\\
   0.059 ~ e^{ -i 1.535} & 0.996 ~ e^{i 0.558} &   0.030 ~ e^{-i 1.364} \\
  0.004 ~ e^{ i 3.055}& 0.018 ~ e^{- i 2.366} &  0.995 ~ e^{ -i 0.199}
\end{array}
\right),
\\
& &
\langle U^{\tilde{g}^\dagger}_4 \rangle \approx
0.881
~ \left(
\begin{array}{ccc}
0.995 ~ e^{-i 0.548}  &     0.004 ~ e^{-2.530} &  0.018 ~ e^{i 1.997} \\
    0.007 ~ e^{ -i 0.375}  &     1.005 ~ e^{i 0.339} & 0.055 ~ e^{i 2.892} \\
    0.030 ~ e^{-i 0.142} &     0.059 ~ e^{i 0.733} &  0.996 ~ e^{i 0.209}
\end{array} 
\right),
\end{eqnarray}
\end{widetext}
where the matrices have been obtained by factoring out the cubic root of the determinant of the link average. This determinant, which is of course the same for the three types of ensembles, is found to be $\det \langle U_4 \rangle = (0.881)^3$.
The matrices have a structure that is essentially diagonal, in agreement with color invariance constraints, and with the phases
adding to zero, in agreement with charge conjugation invariance constraints. More precisely, we find that, for each ensemble, one of the diagonal elements is $\approx e^{i 2\pi/3L_4 k}= e^{i 0.349 k}$ with $k=0,-1$ or $1$, while the product of the other two is $\approx e^{-i 2\pi/3L_4 k}=e^{-i 0.349 k}$. As discussed above, center-symmetry would require that the three diagonal elements are equal to $e^{i 2\pi/3L_4 k}$, with $k=-1,1$ and $0$, no matter what type of ensemble is considered. Thus the high temperature data indicate that center symmetry is broken.

On the other hand, with the low temperature data on the  $64^3\times8$ lattice we find that 
\begin{widetext}

\beq
& & 
\langle U_4 \rangle\approx
0.855
~ \left(
\begin{array}{lll}
0.995 ~ e^{- i 0.230} & 0.068 ~ e^{ i 1.935}& 0.030 ~ e^{ i 2.597}\\
0.053~ e^{i 1.441} &  1.007 ~ e^{i 0.284}& 0.026 ~ e^{- i 1.429}\\
0.040 ~ e^{i 0.778}& 0.031 ~ e^{-i 0.832}& 0.993~ e^{- i 0.058}
\end{array}
\right),
\\
& &
\langle U^{\tilde g}_4 \rangle \approx
0.855
~ \left(
\begin{array}{lll}
   0.993 ~ e^{- i 0.320} &  0.040 ~ e^{i 1.302} & 0.031 ~ e^{- i 2.665} \\
   0.030 ~ e^{i 2.335} & 0.995 ~ e^{i 0.294} & 0.068 ~ e^{i 0.102} \\
   0.026 ~ e^{-i 0.120} & 0.053 ~ e^{-i 2.747} &  1.007 ~ e^{i 0.022} 
\end{array}
\right),
\\
& &
\langle U^{\tilde{g}^\dagger}_4 \rangle\approx
0.855
~ \left(
\begin{array}{l@{\hspace{0.75cm}}l@{\hspace{0.75cm}}l}
1.007 ~ e^{- i 0.240} &  0.026 ~ e^{ i 0.404} & 0.053 ~ e^{i 1.703} \\
0.031~ e^{- i 2.926} &  0.993 ~ e^{ i 0.203} &  
  0.040 ~ e^{- i 0.531} \\
  0.068 ~ e^{i 1.411} &   0.030 ~ e^{- i 1.854} &  0.995 ~ e^{ i 0.032} 
\end{array} 
\right).
\eeq
\end{widetext}
Although the results are not fully conclusive due to statistical errors, the data seem more compatible with an unbroken center symmetry where, no matter what ensemble is considered, the diagonal elements are close to $\approx e^{i 2\pi/3L_4 k}= e^{i 0.262 k}$ with $k=-1,1$ and $0$.

To reduce the numerical errors, we now consider the double average, both over the ensemble and over the lattice sites. For a temperature above the transition, i.e. for the simulation using a $\smash{L_4=6}$, we find

\begin{center}
\begin{figure*}[t]
\includegraphics[height=0.2\textheight]{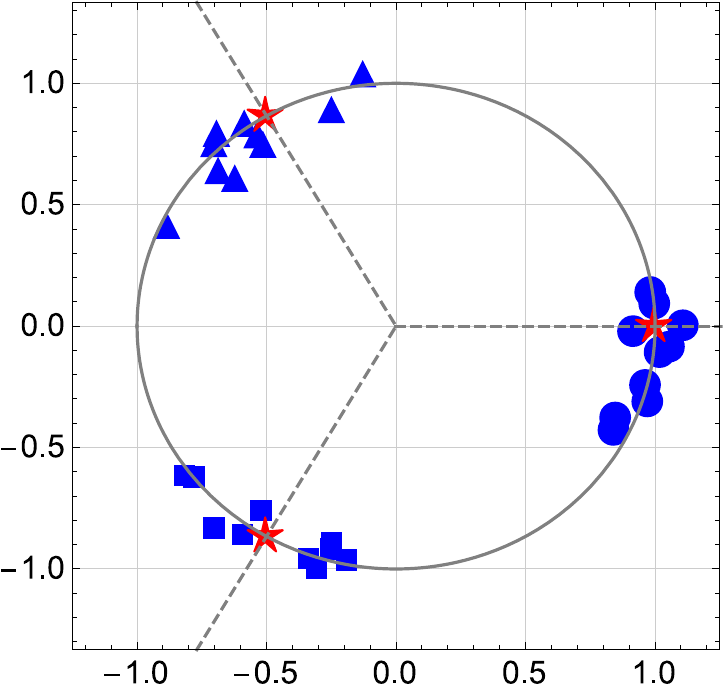}\qquad\includegraphics[height=0.2\textheight]{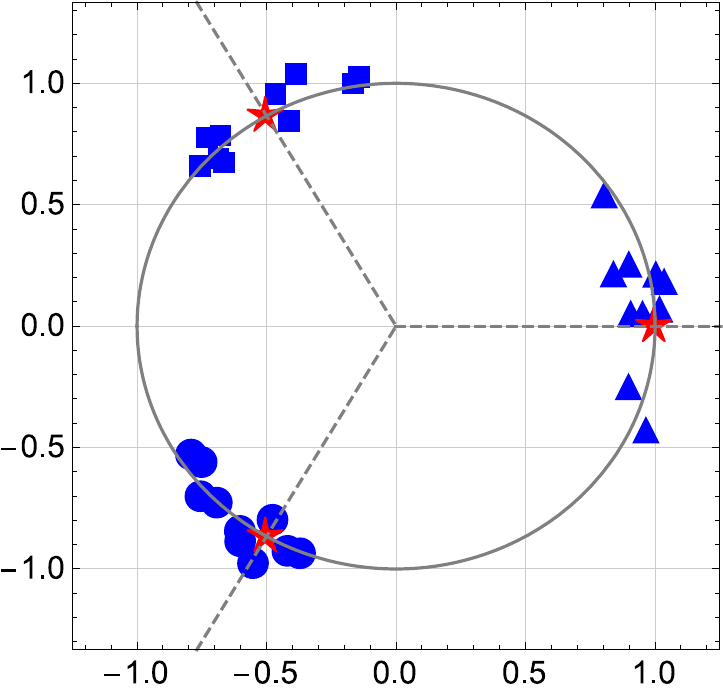}\qquad\includegraphics[height=0.2\textheight]{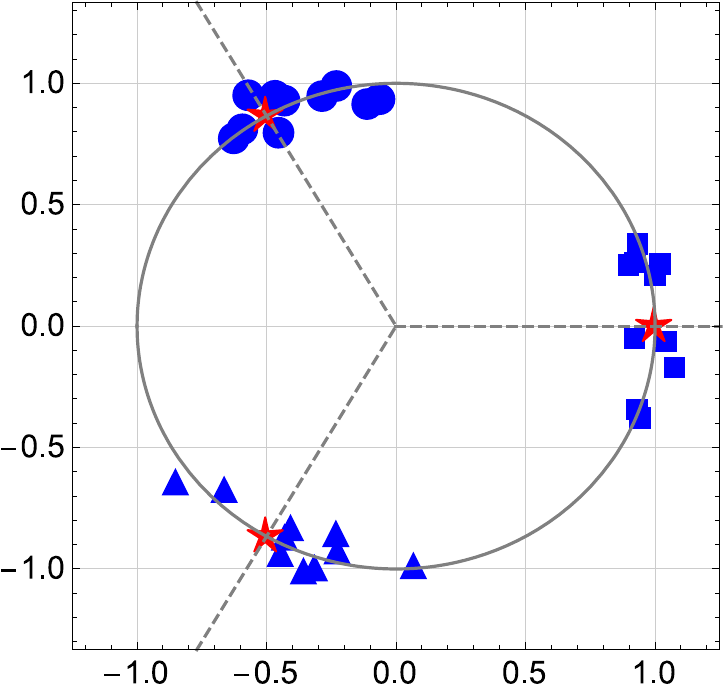}\\
\vglue4mm
\includegraphics[height=0.2\textheight]{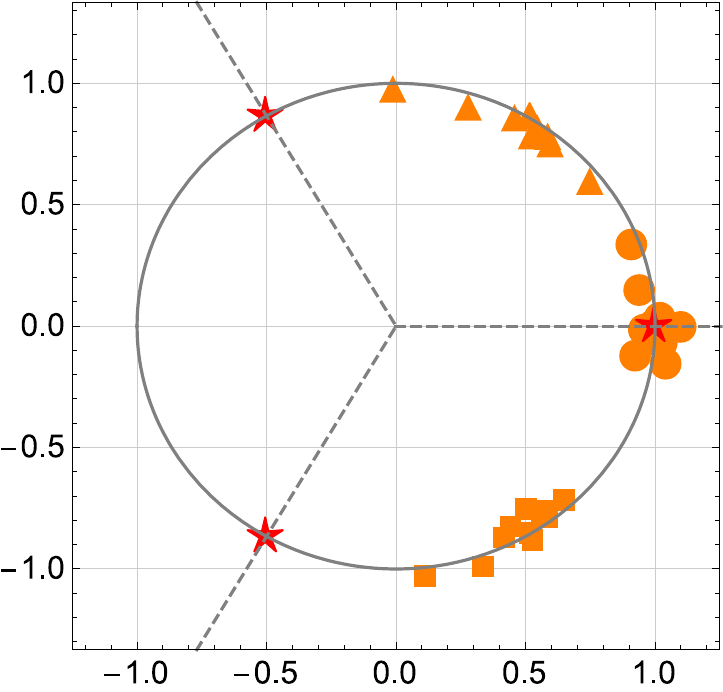}\qquad\includegraphics[height=0.2\textheight]{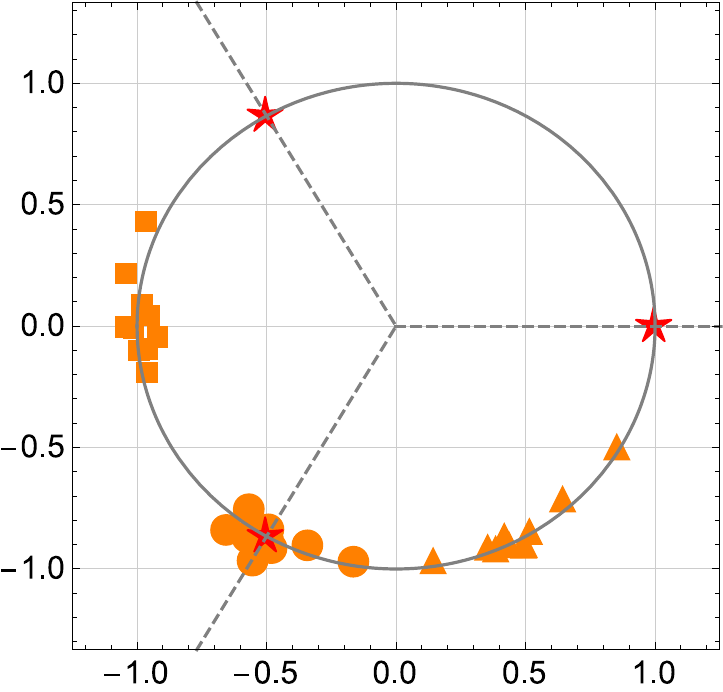}\qquad\includegraphics[height=0.2\textheight]{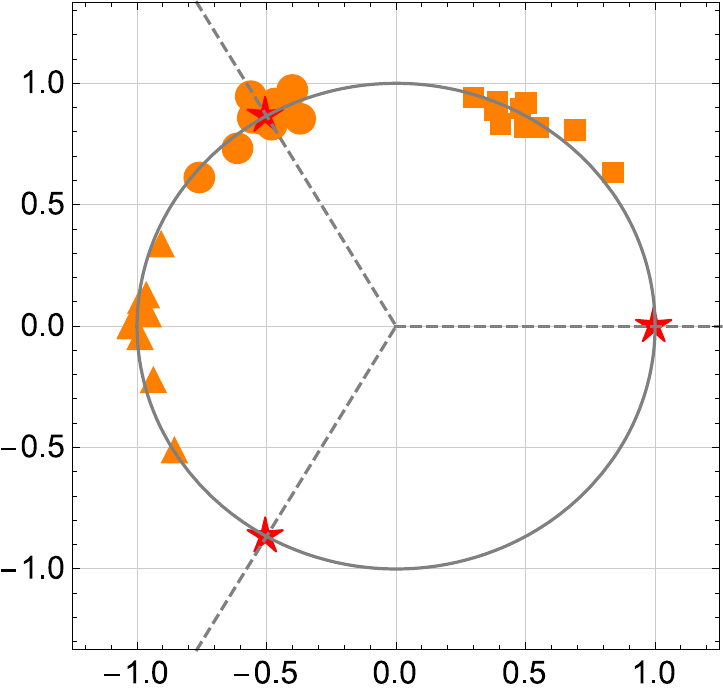}
\caption{Location in the complex plane of the first (square), second (triangles) and third (circles) diagonal elements of $M^{L_4}$ as obtained from averaging over gauge-fixed ensembles of the type $0$ (left), $+$ (middle) or $-$ (right), below (top) and above (bottom) the deconfinement transition, at ten lattice sites.}
\label{fig:diag}
\end{figure*}
\end{center}

\begin{widetext}

\beq
& & 
\langle\langle U_4 \rangle\rangle \approx
0.866
\left(
\begin{array}{lll}
 1.000 ~ e^{- i 0.169} & 0.000 & 0.000 \\
 0.000 & 1.000 ~ e^{i 0.169} &  0.000\\
0.000 &  0.000 & 1.000 ~ e^{ i 0.000} 
\end{array}
\right),
\\
& & 
\langle\langle U^{\tilde g}_4 \rangle\rangle \approx
0.866
\left(
\begin{array}{lll}
1.000 ~ e^{- i 0.349} & 0.000 & 0.000 \\ 
0.000 &  1.000 ~e^{i 0.529} & 0.000 \\
0.000 & 0.000 & 1.000~e^{ -i 0.180}
\end{array}
\right),
\\
& & 
\langle\langle U^{\tilde{g}^\dagger}_4 \rangle\rangle \approx
0.866
\left(
\begin{array}{lll}
 1.000 ~ e^{- i 0.529 } & 0.000 & 0.000 \\
 0.000 &  1.000 ~ e^{i 0.349} & 0.000 \\
 0.000 & 0.000 & 1.000 ~ e^{i 0.180}
\end{array}
\right),
\eeq
whereas, at a temperature below the transition corresponding to $\smash{L_4=8}$, we find 
\beq
& & 
\langle\langle U_4 \rangle\rangle \approx
0.862
~ \left(
\begin{array}{lll}
 1.000 ~ e^{-i 0.257 } & 0.000 & 0.000 \\
 0.000 &  1.000 ~ e^{i 0.257}  &  0.000 \\
 0.000 & 0.000 &  1.000 ~ e^{i 0.000}
\end{array}
\right),
\\
& & 
\langle\langle U^{\tilde g}_4 \rangle\rangle \approx 
0.862
\left(
\begin{array}{lll}
 1.000 ~ e^{-i 0.262} & 0.000 & 0.000 \\
 0.000 & 1.000 ~ e^{i 0.267} & 0.000\\
 0.000 & 0.000 & 1.000 ~ e^{-i 0.005} 
\end{array}
\right),
\\
& & 
\langle\langle U^{\tilde{g}^\dagger}_4 \rangle\rangle \approx
0.862
\left(
\begin{array}{lll}
1.000 ~ e^{- i 0.266}  & 0.000 & 0.000 \\
0.000 & 1.000 ~ e^{ i 0.261}  & 0.000 \\
0.000 & 0.000 & 1.000 ~ e^{ i 0.005} 
\end{array}
\right).
\eeq

\end{widetext}

The main features described above on the case of the lattice origin are now much clearly visible. The matrices are clearly diagonal and follow the constraints of charge conjugation with the phases $e^{ i 0.349}$ and $e^{ i 0.262}$ playing a central role for the high and low temperature data respectively. It is also now pretty clear that the low temperature data is compatible with an unbroken center symmetry, as the diagonal elements are all of the form $e^{ i 0.262k}$, thus reproducing the predicted structure given in Eq. (\ref{eq:1pt}). The high temperature data, in contrast, break the symmetry as two of the diagonal elements do not follow the pattern $e^{ i 0.349k}$. Then, the numerical simulations confirm the theoretical predictions on the use of the average link as an order parameter for the transition from the confined to deconfined phase.

The particular role of the phases $\exp\{-i/L_4 2\pi/3\}$ and $\exp\{i/L_4 2\pi/3\}$ can be nicely illustrated by plotting the distances $|M_{11}-\exp\{-i/\hat{L}_4 2\pi/3\}|$ and $|M_{22}-\exp\{i/\hat{L}_4 2\pi/3\}|$, where $\smash{M\equiv \langle U_4\rangle/\eta}$, as functions of a  control parameter $\hat{L}_4$ whose physical value is the temporal extent $L_4$ of the lattice ($8$ for the low temperature data and $6$ for the high temperature data), see Fig.~\ref{fig:dist}. The error bars result from the propagation of the error from the link average to the above-defined distances. At low temperature, we see that the distances are small (and essentially the smallest) when $\hat L_4$ takes its physical value (vertical blue line), but this is not true in the high temperature phase. This was done for the origin of the lattice (left plot) and we have checked that this pattern extends to the other sites of the lattice. In the right plot of Fig.~\ref{fig:dist}, we show the corresponding test of (\ref{eq:38}) upon averaging over ten sites to reduce statistical fluctuations.

Finally, in Fig.~\ref{fig:diag}, we show the location of the three diagonal elements of $M$, or more conveniently $M^{L_4}$,\footnote{In particular, note that the center symmetry constraint on $M^{L_4}$ is the one that remains non-trivial in the continuum limit (corresponding to $\smash{L_4\to\infty}$ at fixed temperature).} in the complex plane below and above the deconfinement transition, when averaging over three different gauge-fixed ensembles, ${\cal E}_{0{\rm gf}}$, ${\cal E}_{+{\rm gf}}$ and ${\cal E}_{-{\rm gf}}$, see the discussion at the end of Sec.~\ref{sec:other}. In the confined phase, these three ensembles are essentially the same and we see that the diagonal elements of $M$ comply with the center symmetry constraints (\ref{eq:38}). In the deconfined phase, the constraints are violated in three different ways, depending on which ensemble is considered.

In line with the discussion above, we find that upon breaking the symmetry, one of the diagonal elements lies along $e^{ik 2\pi/3}$, with $k=0,-1$ or $1$, while the product of the other two has a phase fixed to $e^{-ik 2\pi/3}$. 

Surprisingly, not only does one element remain along $e^{ik 2\pi/3}$ but it is actually of modulus $1$. We have found no explanation of this fact but, given this observation, we find that the matrix $M$ is unitary. This is not obvious a priori since an average of unitary matrices such as $\langle U_\mu\rangle$ is not necessarily unitary. What we know from the very definition of $M$ is that $\smash{{\rm det}\,M =1}$

In the confined phase, the unitarity of $M$ follows from the center symmetry constraint (\ref{eq:38}). In the deconfined phase, we need to invoke color invariance which ensures that $M$ is diagonal and charge conjugation which ensures that two of the diagonal elements have the same modulus $\rho$ while their product has a phase that cancels that of the third element. Then, one can write $\smash{1={\rm det}\,M=\rho^2}$, where we have used that the third element has modulus $1$ as we have empirically observed in the data. It follows that $\smash{\rho=1}$ and thus that $\smash{|M_{11}|=|M_{22}|=|M_{33}|=1}$. This implies that $M$ is unitary.

\section{Additional remarks}\label{sec:add}

\subsection{Twisted links}
To make the center-symmetry constraints look simpler, in particular in view of a future discussion of general link correlators, it is useful to work in terms of twisted link variables. First, it is convenient to see $g_c(\hat\mu)$ defined in  Eq.~(\ref{eq:13}) as deriving from a more general function valid on all sites $n$ of the lattice:
\beq
g_c(n)\equiv e^{-i\frac{n_4}{L_4}\frac{4\pi}{3}\frac{\lambda_3}{2}}\,,\label{eq:gcn}
\eeq
and not just on the first sites $\smash{n=\hat\mu}$ next to the origin along each direction $\mu$. We then notice that
\beq
g^\dagger_c(\hat\mu)=g_c^\dagger(n+\hat\mu)g_c(n)\,,
\eeq
and
\beq
\tilde F[U] & = & \sum_{n,\mu} {\rm Re}\, {\rm tr}\, g_c^\dagger(n+\hat\mu)g_c(n) U_\mu(n)\nonumber\\
& = & \sum_{n,\mu} {\rm Re}\, {\rm tr}\, g_c(n) U_\mu(n)g_c^\dagger(n+\hat\mu)\nonumber\\
& = & \sum_{n,\mu} {\rm Re}\, {\rm tr}\, U^{g_c}_\mu(n)\,.
\eeq
which looks like the standard Landau gauge functional, but for the transformed link  
variable $U_\mu^{g_c}$, which we refer as a twisted link in what follows. Note that $g_c(n)$ is not periodic, not even periodic modulo a center element, so the twisted links are not periodic along the temporal directions. But one can construct  correlators of twisted links from which it is very simple to retrieve (if needed) the correlators of the original links.

\begin{center}
\begin{figure}[t]
\includegraphics[height=0.3\textheight]{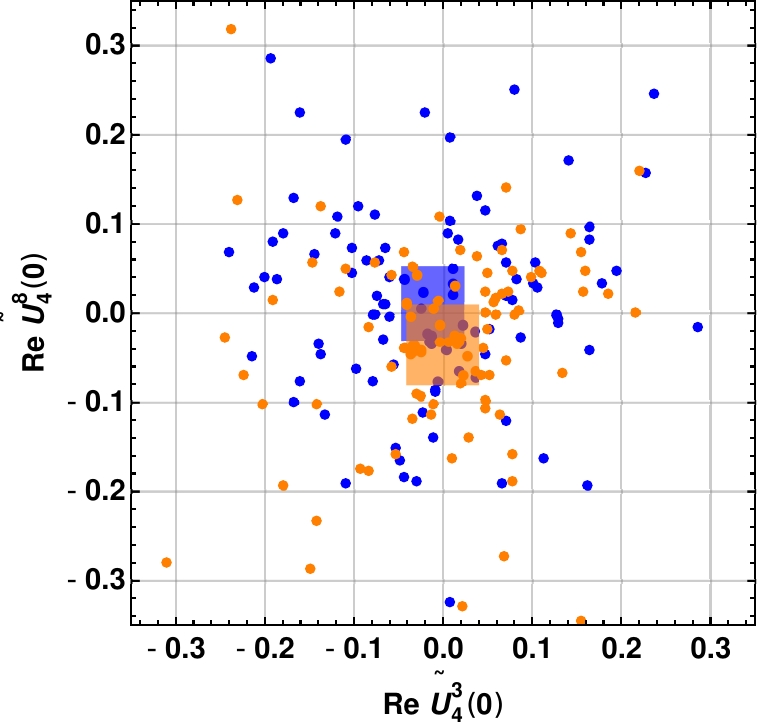}\\
\vglue4mm
\includegraphics[height=0.3\textheight]{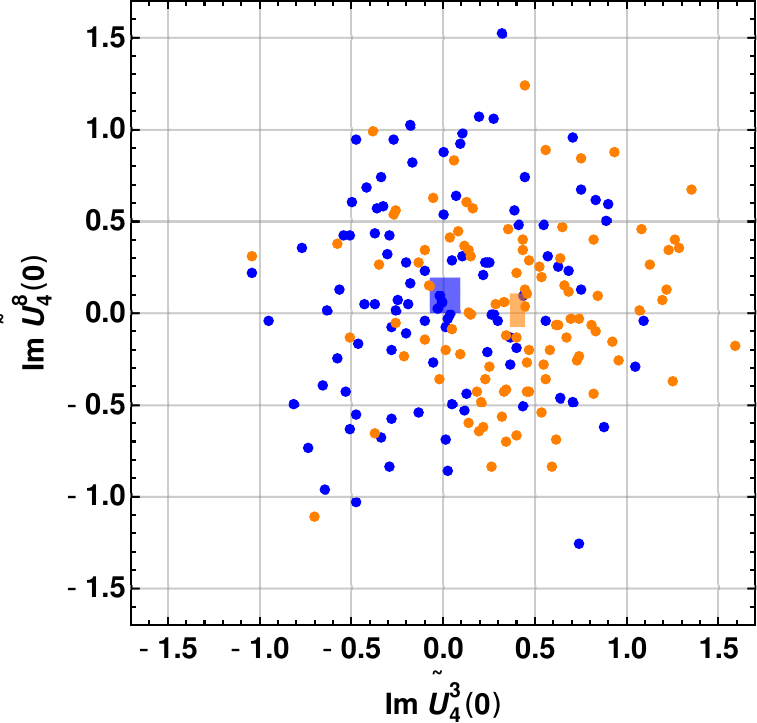}
\caption{Spread of the real and imaginary parts of $\hat U_4^3(n=0)$ and $\hat U_4^8(n=0)$ below (blue) and above (orange) the transition temperature. The respective mean values and standard deviations are represented by the shaded rectangles.}
\label{fig:re_im}
\end{figure}
\end{center}

Now, since 
\beq
(U_\mu^{g_0})^{g_c}=U_\mu^{g_cg_0}=U_\mu^{g_cg_0g_c^\dagger g_c}=(U_\mu^{g_c})^{g_c g_0 g_c^\dagger}\,,
\eeq
the maximization of the functional $\tilde F[U]$ over the ${\cal G}_0$-orbit of $U_\mu$ is equivalent to the maximization of the standard Landau gauge functional $\sum_{n,\mu} {\rm Re}\, {\rm tr}\, U_\mu(n)$ over the $g_c{\cal G}_0g_c^\dagger$-orbit of $U_\mu^{g_c}$. The elements of this orbit are all twisted links, that is of the form $U_\mu^{g_c}$ with $U_\mu$ periodic. It follows that the gauge fixing associated to the gauge $\tilde F[U]$ can be interpreted as a gauge fixing associated to the standard Landau gauge, to the price of working with twisted links (obeying modified, non-periodic, boundary conditions) and to replacing ${\cal G}_0$ by $g_c{\cal G}_0g_c^\dagger$ which is different from ${\cal G}_0$.

An interesting property of the twisted links is that they transform in a simpler way than the standard links. Indeed under a center transformation (\ref{eq:V_transfo}), one has
\beq
& & U^{g_c}_\mu(n)\nonumber\\
& & \to\,g_c(n)\tilde g(n)g_c^\dagger(n)U^{g_c}_\mu(n)g_c(n+\hat\mu)\tilde g^\dagger(n+\hat\mu)g^\dagger_c(n+\hat\mu)\,.\nonumber\\
\eeq
But from Eq.~(\ref{eq:26}), we have
\beq
& & g_c(n+\hat\mu)\tilde g^\dagger(n+\hat\mu)g_c^\dagger(n+\hat\mu)\nonumber\\
& & \hspace{0.5cm}=\,g_c(n)g_c(\hat\mu)\tilde g^\dagger(n+\hat\mu)g_c^\dagger(n+\hat\mu)\nonumber\\
& & \hspace{0.5cm}=\,g_c(n)\tilde g^\dagger(n)g_c(\hat\mu)g_c^\dagger(n+\hat\mu)\nonumber\\
& & \hspace{0.5cm}=\,g_c(n)\tilde g^\dagger(n)g_c^\dagger(n)\,.
\eeq
In other words, at each lattice site $n$, the twisted link transforms under a local color rotation
\beq
& & g_c(n)\tilde g(n)g_c^\dagger(n)\nonumber\\
& & \hspace{0.5cm}=e^{-i\frac{n_4}{L_4}\frac{4\pi}{3}\frac{\lambda_3}{2}}W_{\alpha_{13}}W_{\alpha_{12}}e^{-i\frac{n_4}{L_4}4\pi\rho_1^jt^j}e^{i\frac{n_4}{L_4}\frac{4\pi}{3}\frac{\lambda_3}{2}}\,.\nonumber\\
\eeq
For later use, let us also mention that, in terms of twisted links, charge conjugation reads
\beq
& & U^{g_c}_\mu(n)\nonumber\\
& & \to\,g_c(n)e^{i\pi\frac{\lambda_1}{2}}g_c(n) (U^{g_c}_\mu)^*(n)g_c^\dagger(n+\hat\mu)e^{-i\pi\frac{\lambda_1}{2}}g_c^\dagger(n+\hat\mu)\,.\nonumber\\
\eeq
Using Eq.~(\ref{eq:CC}), this becomes
\beq
U^{g_c}_\mu(n)\to e^{i\pi\frac{\lambda_1}{2}} (U_\mu^{g_c}(n))^*e^{-i\pi\frac{\lambda_1}{2}}\,.
\eeq

\begin{center}
\begin{figure}[t]
\includegraphics[height=0.3\textheight]{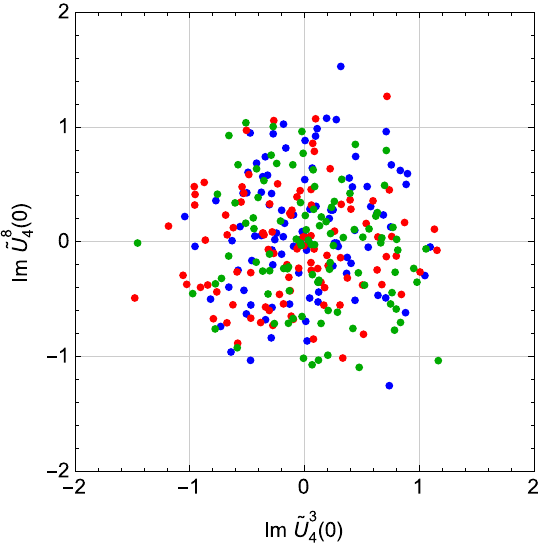}\\
\vglue4mm
\includegraphics[height=0.3\textheight]{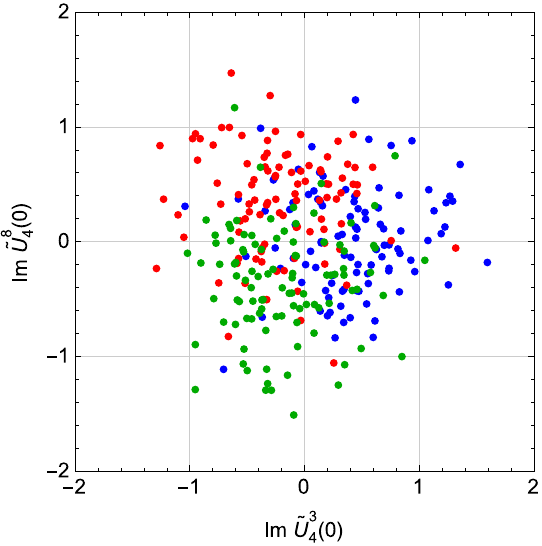}
\caption{The three gauge-fixed ensembles ${\cal E}_{0,{\rm gf}}$ (blue), $\smash{{\cal E}_{-,{\rm gf}}={\cal E}_{0,{\rm gf}}^g}$ (red) and $\smash{{\cal E}_{+,{\rm gf}}={\cal E}_{0,{\rm gf}}^{g^\dagger}}$ (green), projected along ${\rm Im}\,\hat U_4^3(0)$ and ${\rm Im}\, \hat U_4^8(0)$, together with their transformation under $g$ and $g^\dagger$. Top: below the transition; Bottom: above the transition.}
\label{fig:sectors}
\end{figure}
\end{center}

\vglue-6mm

Denoting the twisted links by $\hat U_\mu(n)$ and $\smash{\hat g(n)\equiv g_c(n)\tilde g(n)g^\dagger_c(n)}$, we have this time
\beq
\langle \hat U_\mu(n)\rangle=\hat g(n)\,\langle \hat U_\mu(n)\rangle\,\hat g^\dagger(n)\,.
\eeq
Restricting to the diagonal components $|\rho\rangle\langle\rho|$ in the basis $|\rho'\rangle\langle\rho|$ (recall that the other components vanish from color invariance), one finds
\beq
\langle \hat U^{\rho\rho}_{\mu}(n)\rangle=\langle \hat U^{{\cal R}\rho {\cal R}\rho}_{\mu}(n)\rangle\,,\label{eq:correlators}
\eeq
where ${\cal R}$ is the $2\pi/3$-rotation defined earlier, and thus
\beq
\langle \hat U^{\rho_1\rho_1}_{\mu}(n)\rangle=\langle \hat U^{\rho_2\rho_2}_{\mu}(n)\rangle=\langle \hat U^{\rho_3\rho_3}_{\mu}(n)\rangle
\eeq
which implies
\beq
\langle \hat U_{\mu}(n)\rangle=\eta\mathds{1}\,.\label{eq:tconstraint}
\eeq
Of course, this can also be directly deduced from (\ref{eq:eta_gc}).

We can now switch to the other basis using Eq.~(\ref{eq:change}). Multiplying each side of Eq.~(\ref{eq:correlators}) by $\rho^i$'s and summing over weights, we find
\beq
\langle \hat U^{j}_{\mu}(n)\rangle & = & \sum_{\rho}\rho^{j}\,\left\langle \hat U^{{\cal R}\rho {\cal R}\rho}_{\mu}(n)\right\rangle\nonumber\\
& = & \sum_{\rho}({\cal R}^{-1}\rho)^{j}\,\langle \hat U^{\rho\rho}_{\mu}(n)\rangle\nonumber\\
& = & ({\cal R}^{-1})^j_{\,\,k}\sum_{\rho}\rho^{k}\,\langle \hat U^{\rho\rho}_{\mu}(n)\rangle\nonumber\\
& = & ({\cal R}^{-1})^j_{\,\,k}\,\langle \hat U^{k}_{\mu}(n)\rangle\,.\label{eq:crrii}
\eeq
This means that the vector $(\langle \hat U^3_{\mu}(n)\rangle,\langle \hat U^8_{\mu}(n)\rangle)$ is invariant under internal rotations of the indices by an angle $-2\pi/3$, and thus that
\beq
\langle \hat U^3_{\mu}(n)\rangle=\langle \hat U^8_{\mu}(n)\rangle=0\,,
\eeq
if center symmetry is not broken. We also note that, for a $\hat C$-invariant ensemble, Eqs.~(\ref{eq:c12}) and (\ref{eq:c3}) combined with Eq.~(\ref{eq:change}) imply that $\langle \hat U^8_{\mu}(n)\rangle$ is real while $\langle \hat U^3_{\mu}(n)\rangle$ is imaginary.\footnote{For the other possible ensembles, the constraints are different. In particular one finds that ${\rm Im}\,\langle \hat U^8_{\mu}(n)\rangle=\pm \sqrt{3}{\rm Im}\,\langle \hat U^3_{\mu}(n)\rangle$, in line with Fig.~\ref{fig:sectors}.} So, the breaking of center symmetry should be visible by monitoring ${\rm Im}\,\langle \hat U^3_{\mu}(n)\rangle$ or ${\rm Re}\,\langle \hat U^8_{\mu}(n)\rangle$.

In Fig.~\ref{fig:re_im}, we evaluate the vectors $\smash{({\rm Re}\,\hat U_4^3,{\rm Re}\,\hat U_4^8)}$ and $\smash{({\rm Im}\,\hat U_4^3,{\rm Im}\,\hat U_4^8)}$ at site $\smash{n=0}$ for all the configurations of our $\hat C$-invariant ensembles ${\cal E}_{0{\rm gf}}$ below and above the transition temperature.\footnote{We show the real part (resp. the imaginary parts) together on the same plot because these are the components that are related to each other by the center symmetry, according to Eq.~(\ref{eq:crrii}).} We see that, below the transition, all these quantities fluctuate around $0$ within errors, in agreement with the center-symmetry constraints. Above the transition, one of these quantities, $\smash{{\rm Im}\, \hat U_4^3(n=0)}$, starts developing fluctuations around a non-zero value, signaling the breaking of center symmetry. We find that ${\rm Re}\,\hat U_4^8(n=0)$ seems to keep fluctuating around $0$, but it could be that it fluctuates around a value which is negligible with respect to the mean value of ${\rm Im}\,\hat U_4^3(n=0)$.

This order parameter pattern can be nicely visualized by looking at the link averages over the three possible ensembles ${\cal E}_{0,{\rm gf}}$, $\smash{{\cal E}_{-,{\rm gf}}={\cal E}_{0,{\rm gf}}^{\tilde g}}$ and $\smash{{\cal E}_{+,{\rm gf}}={\cal E}_{0,{\rm gf}}^{\tilde g^\dagger}}$, see Fig.~\ref{fig:sectors}. In the symmetric phase, the system explores enough phase-space, and the three ensemble are essentially the same, see the top plot of Fig.~\ref{fig:sectors}.

In  the broken phase, the system gets stuck in a given sector characterized by the value of the vector $({\rm Im}\,\langle U_4^3\rangle, {\rm Im}\,\langle U_4^8\rangle)$,\footnote{It may seem strange that  ${\rm Im}\,\langle U_4^8\rangle$ takes a non-zero value since we said above that $\langle U_4^8\rangle$ should be real from charge conjugation invariance. However, as we already pointed out, in the broken phase, this statement applies only to the average over the $\hat C$-invariant ensemble.} see the bottom plot of Fig.~\ref{fig:sectors}.\\

\subsection{Mapping the gluon field}
So far, we have considered link averages. However, in the continuum theory, we work with gauge fields.
It is thus interesting to look at the gauge field average. A non-trivial check is whether one retrieves the center symmetry constraint for the gluon field average
\beq
\beta_T\langle A_4^3(x)\rangle=-\frac{4\pi}{3}\,
\eeq
with $\smash{\beta_T\equiv 1/T}$,
that is expected in the continuum, see the Appendix.

The links are path ordered exponentiations of the gluon fields and, for sufficiently small lattice spacing, one can write
\beq
U_4(n)\!=\! e^{ia A_4(n')}\,,\label{eq:start}
\eeq
where $\smash{A_4(n')\equiv A^a_4(n')t^a}$ and $n^\prime$ is a point between $n$ and $n + \hat{4}$, where $\hat{4}$ is the unit vector along Euclidean time direction. As usual,
the exponential is defined via its Taylor series, that is an expansion in $a A$ and, typically, the terms are labeled by the corresponding power
of the lattice spacing. 

Let us first consider this expansion to linear order in $aA$:
\beq
U_4(n)=\mathds{1}\!+\!ia A_4(n')+{\cal O}(a^2)\,.\label{eq:formula}
\eeq
Upon taking the dagger of this expression, one finds
\beq
U^\dagger_4(n)=\mathds{1}\!-\!ia A_4(n')+{\cal O}(a^2)\,,
\eeq
and because the ${\cal O}(a^2)$ terms are hermitian, one gets
\beq
U_4(n)-U^\dagger_4(n)=2ia A_4(n')+{\cal O}(a^3)\,.\label{eq:UA}
\eeq
This means that the quantity
\beq
aA^{\rm LO}_4(n')\equiv\frac{U_4(n)-U^\dagger_4(n)}{2i}\label{eq:LO_est}
\eeq
provides an estimate of $aA$ to $a^3$ accuracy:
\beq
aA^{\rm LO}_4(n')=a A_4(n')+{\cal O}(a^3)\,,
\eeq
from which one can evaluate the average  of $aA$ to $a^3$ accuracy:
\beq
\langle aA^{\rm LO}_4(n')\rangle=\langle a A_4(n')\rangle+{\cal O}(a^3)\,.\label{eq:vevUA}
\eeq
and also the double average. For the latter, using this estimate on our low temperature data, we find
\footnote{Statistical errors computed with the jackknife method.}
\beq
\beta_T\langle\langle A_4^{\rm LO,3}\rangle\rangle= -3.5079(60)       \label{eq:LO}
\eeq
which is not exactly close to the expected value $-4\pi/3\simeq -4.18879$.

One possible culprit is that $aA$ might be not small enough for the higher order terms in Eq.~(\ref{eq:formula}) to be negligible. One simple way to illustrate this is to notice that Eq.~(\ref{eq:formula}) implies
\beq
{\rm det}\,\langle U_4(n)\rangle=1+{\cal O}(a^2)\,,\label{eq:det}
\eeq
where we used that ${\rm det}(\mathds{1}+\delta M)\simeq 1+{\rm tr}\,\delta M$ as well as ${\rm tr}\,t^a=0$. Therefore, even though the average of a link is not a unitary matrix, its determinant should approach $1$ in the continuum limit, at least according to (\ref{eq:formula}). Our results in the previous section show, however, that ${\rm det}\,\langle U_4(n)\rangle$ deviates from $1$ by $36\%$ both in the confined and deconfined phases. This indeed indicates that higher order terms in Eq.~(\ref{eq:formula}) should be considered. And in fact, the next-to-leading term in (\ref{eq:det}) can be expressed in terms of the connected two-point correlator
\beq
{\rm det}\,\langle U_4(n)\rangle=1-\frac{a^2}{2}{\rm tr}\,\Big\langle \Big(A-\langle A\rangle\Big)^2\Big\rangle+{\cal O}(a^4)\,,\label{eq:det}
\eeq
and represents a negative contribution. It would be interesting to evaluate this contribution and compare it to the observed deviation but we leave for a future investigation where we will have a closer look at the propagator.

We have considered higher order terms in this formula that correct Eq.~(\ref{eq:UA}) and thus our determination of the gluon field from the links. Let us illustrate how this works to obtain a determination of $aA$ from the links valid to $a^5$ accuracy. The trick is to write Eq.~(\ref{eq:UA}) to next order:
\beq
& & U_4(n)-U^\dagger_4(n)\nonumber\\
& & \hspace{0.5cm}=\,2ia A_4(n')-\frac{1}{3}i(aA_4(n'))^3\!\!+\!{\cal O}(a^5)\,,
\eeq
and a similar expansion for $U_4^2-(U_4^\dagger)^2$
\beq
& & (U_4(n))^2-(U^\dagger_4(n))^2\nonumber\\
& & \hspace{0.5cm}=\,4ia A_4(n')-\frac{8}{3}i(aA_4(n'))^3\!\!+\!{\cal O}(a^5)\,,
\eeq
and to find a linear combination of the two that kills the terms of order $a^3$. This linear combination is easily found to be
\beq
& &8(U_4(n)-U^\dagger_4(n))-((U_4(n))^2-(U^\dagger_4(n))^2)\nonumber\\
& & \hspace{1.0cm}= 12ia A_4(n')+{\cal O}(a^5)\,.
\eeq
It follows that
\beq
aA^{\rm NLO}_4(n') & \equiv & \frac{2}{3}\frac{U_4(n)-U^\dagger_4(n)}{i}\nonumber\\
& - & \frac{1}{12}\frac{((U_4(n))^2-(U^\dagger_4(n))^2}{i}
\eeq
provides an estimate of $aA$ to $a^5$ accuracy:
\beq
aA^{\rm NLO}_4(n')=a A_4(n')+{\cal O}(a^5)\,,
\eeq
from which one can evaluate the average of $aA$ to $a^5$ accuracy:
\beq
\langle aA^{\rm NLO}_4(n')\rangle =\langle a A_4(n')\rangle+{\cal O}(a^5)\,.
\eeq
Similarly, one can construct an estimator of $aA$ to $a^7$ accuracy that also involves $U_4^3$
\beq
aA^{\rm NNLO}_4(n') & \equiv & \frac{3}{4}\frac{U_4(n)-U^\dagger_4(n)}{i}\nonumber\\
& - & \frac{3}{20}\frac{((U_4(n))^2-(U^\dagger_4(n))^2}{i}\nonumber\\
& + & \frac{1}{60}\frac{((U_4(n))^3-(U^\dagger_4(n))^3}{i}\,.
\eeq

Using these estimators on our low temperature data, we find
\beq
\beta_T\langle\langle A_4^{\rm NLO,3}\rangle\rangle = -3.8291(67) \label{eq:NLO}
\eeq 
and
\beq
\beta_T\langle\langle A_4^{\rm NNLO,3}\rangle\rangle = -3.8741(68) \label{eq:NNLO}
\eeq
\vglue1mm
\noindent{respectively, which improve over the leading order estimate (\ref{eq:LO}) and seem to converge to the expected continuum value.}

Let us mention that the deviation (\ref{eq:LO}) that we observed at leading order could have been anticipated using the center symmetry constraints on the link average and the measured value of $\eta\neq 1$. Indeed, one can plug (\ref{eq:1pt}) in the RHS of Eq. (\ref{eq:LO_est}), multiply by $L_4$ and use $a L_4=\beta_T$ to arrive at
\beq
\beta_T\langle\langle A_4^{\rm LO,3}\rangle\rangle= -\eta\frac{4\pi}{3}\,.
\eeq
With the value of $\eta\simeq 0.862$ found in the previous section, we arrive at
  \beq
\beta_T\langle\langle A_4^{\rm LO,3}\rangle\rangle\simeq -3.61  \,,
\eeq
not far from the leading order result (\ref{eq:LO}) quoted above. It would be interesting to repeat the same exercise for the improved formulas relating the link to the gauge field and see if one gets indeed a result closer to $-4\pi/3$ as obtained in Eqs. (\ref{eq:NLO}) and (\ref{eq:NNLO}).\footnote{This would require deriving the symmetry constraints for objects such as $\langle U^2\rangle$ or $\langle U^3\rangle$.} Also, it would be interesting to compute the gluon field from the twisted links, since in that case we have $\beta_T\langle \tilde{A}_4^{3}(x)\rangle=0$. Furthermore, one also needs to study the continuum limit by performing simulations with smaller lattice spacings.  Finally, another possibility is that the relation (\ref{eq:start}) needs to be revisited.\footnote{Before gauge-fixing, the functional integral is dominated by pure gauge configurations of the form $g_0(n)g_0^\dagger(n+\hat\mu)$, where $g_0(n)$ is arbitrary and has no reason to become smooth in the continuum. Close to the continuum limit, the links that dominate are then of the form $U_\mu(n)=g_0(n)e^{ia A^a_\mu(x+(a/2)\hat{\mu})t^a}g_0^\dagger(n+\hat\mu)$ but because $g_0(n)$ is not necessarily smooth, some of these links do not approach $\mathds{1}$ in the continuum.
As long as, one remains at a non-gauge-fixed level and evaluates observables such that ${\cal O}[U^{g_0}]={\cal O}[U]$, this is not a problem because the observables do not see the gauge transformation $g_0(n)$ and one can do as if all links that dominate where of the form (\ref{eq:formula}). In contrast, within a generic gauge-fixed setting, the configurations of the gauge-fixed ensemble remain a priori of the general form above, unless the considered gauge and the selection of copies lead to smooth enough $g_0$'s. It would require a separate study, far beyond the scope of the present one, to analyze whether this actually happens in the Landau gauge or in the present center-symmetric Landau gauge.} We leave these interesting questions for future work.

Our study shows that higher terms in the relation between the link and the gluon field have a sizable effect on the outcome and should not be ignored. In general the inclusion of the higher order terms is not a trivial exercise. A particular  example is the evaluation of the plaquette in lattice perturbation theory in comparison to the outcome of Monte Carlo simulations that shows a sizable difference, oftentimes interpreted as due to a nonperturbative contribution that is associated with the gluon condensate $\langle G^2 \rangle$,
see e.g. \cite{Shifman:1978bx,Martinelli:1996pk,Horsley:2012ra} and reference therein.

\subsection{Gribov copies}

In this section, we study the transformation properties of our center-symmetric Landau gauge-fixed ensemble under $\tilde g$. The diagonal components of the links transform according to Eqs.~(\ref{eq:69})-(\ref{eq:71}). This means that to test whether the center-symmetry Landau gauge-fixed ensemble is approximately invariant under $\tilde g$, we can compare $U_4^{\rho_1\rho_1}(n)$ and $e^{-\frac{i}{L_4}\frac{4\pi}{3}} U_4^{\rho_2\rho_2}(n)$ on the available configurations. This comparison is shown in the upper plots of Fig.~\ref{fig:copies}, together with a similar comparison for the Landau gauge configurations in the lower plots, in which case we do not expect the gauge-fixed ensemble to be invariant under $g^\dagger$. One clearly sees a difference between the configurations in those two gauges. In the former case, the sets $U_4^{\rho_1\rho_1}(n)$ and $e^{-\frac{i}{L_4}\frac{4\pi}{3}} U_4^{\rho_2\rho_2}(n)$ are quite different from each other, both at low and at high temperatures. In the case of the center-symmetric Landau gauge in contrast, the sets look pretty much on top of each other at low temperatures, and start deviating slightly only at at high temperatures.

\begin{center}
\begin{figure}[t]
\includegraphics[height=0.2\textheight]{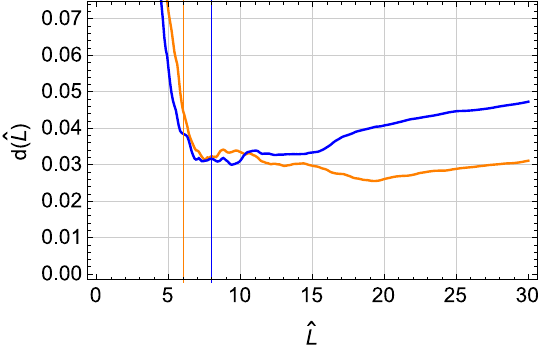}\\
\vglue2mm
\includegraphics[height=0.2\textheight]{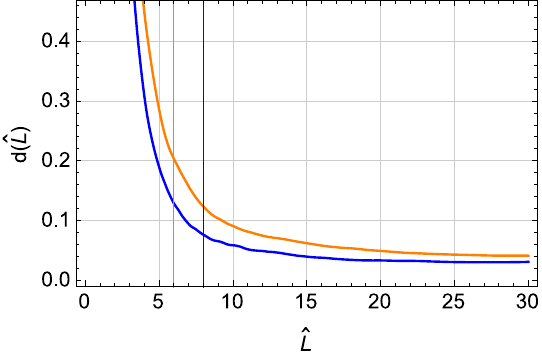}\\
\caption{The distance between the sets $U^{\rho_1\rho_1}_4(0)$ and $e^{-\frac{i}{\hat{L}_4}\frac{4\pi}{3}}U^{\rho_2\rho_2}_4(0)$ as a function of the control parameter $\hat L_4$, at low (blue) and high (orange) temperatures. Top  plot: center-symmetric Landau gauge; Bottom plot: Landau gauge. The vertical lines represent the lattice values of the parameter $\hat L_4$ ($8$ at low temperature and $6$ at high temperature).}
\label{fig:copies_dist}
\end{figure}
\end{center}

The same results can be illustrated by computing the distance\footnote{Given two sets $Z_1$ and $Z_2$ of $N$ points each in the complex, we define their relative distance as $$\frac{1}{N}\sum_{z_1\in Z_1} {\rm  Min}_{z_2\in Z_2}|z_1-z_2|\,.$$} between the sets $U_4^{\rho_1\rho_1}(n)$ and $e^{-\frac{i}{\hat L_4}\frac{4\pi}{3}}U_4^{\rho_2\rho_2}(n)$ with $\hat L_4$ some  auxiliary variable, both in the center-symmetric Landau gauge and in the standard
\begin{widetext}
\begin{center}
\begin{figure}[t]
\includegraphics[height=0.28\textheight]{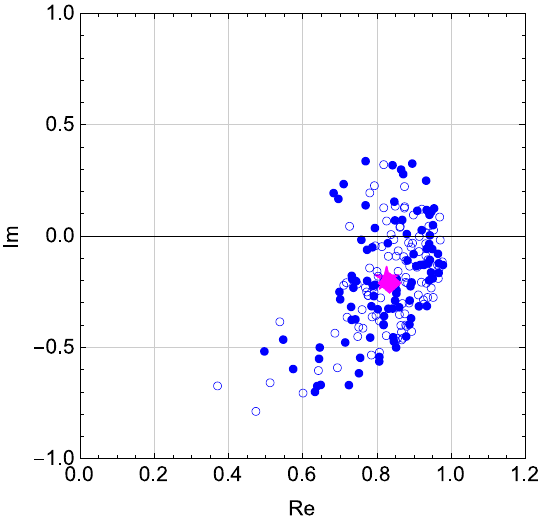}\qquad\includegraphics[height=0.28\textheight]{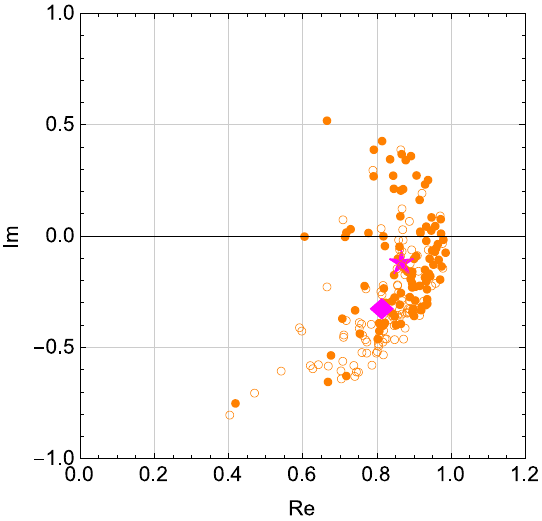}\\
\vglue4mm
\includegraphics[height=0.28\textheight]{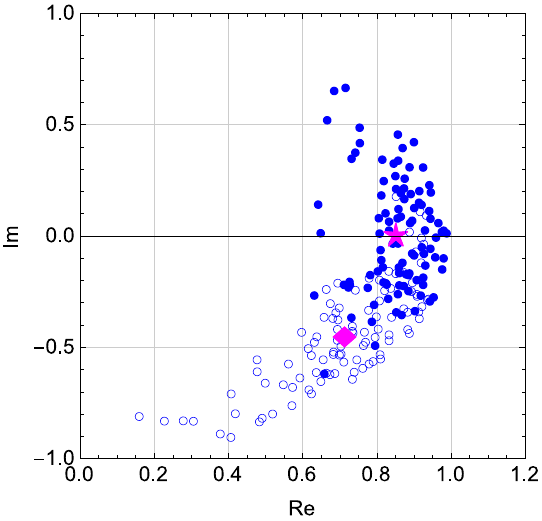}\qquad\includegraphics[height=0.28\textheight]{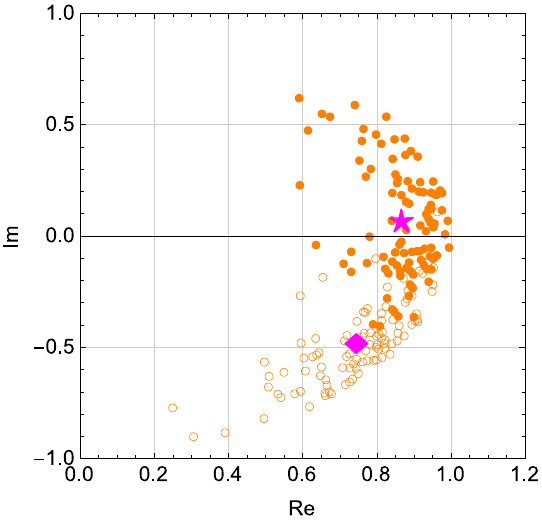}
\caption{The real and imaginary parts of the configurations $U^{\rho_1\rho_1}_4(0)$ (plain circles) and $e^{-\frac{i}{L_4}\frac{4\pi}{3}}U^{\rho_2\rho_2}_4(0)$ (empty circles), at low (blue, left) and high (orange, right) temperatures. The purple star and purple diamond represent the corresponding configuration averages. Top plot: center-symmetric Landau gauge configurations; Bottom plot: Landau gauge configurations.}
\label{fig:copies}
\end{figure}
\end{center}
\end{widetext}
 Landau gauge, see Fig.~\ref{fig:copies_dist}. One clearly sees that nothing peculiar happens in the Landau gauge when $\hat L_4$ takes the lattice value. In contrast, in the center-symmetric Landau gauge, the distance between the two sets is minimized at low temperatures around $\smash{\hat L_4=8}$, the appropriate lattice value for the low temperature data. In the case of the high temperature data, the distance is minimized for a value of $\hat L_4$, that has nothing to do with the lattice value $\smash{\hat L_4=6}$.

These results illustrate that despite the presence of Gribov copies, our gauge-fixing procedure produces an approximately $\tilde g$-invariant ensemble in the symmetric phase. Note also that our center-symmetric Landau gauge-fixed ensemble was produced by starting from an already available Landau gauge-fixed ensemble. That the final gauge-fixed ensemble turns out to be approximately $\tilde g$-invariant seems to be in line with the discussion at the end of Sec.~\ref{sec:Gribov}.

\section{Conclusions}

In this work, we have considered the lattice implementation of gauge-fixed pure Yang-Mills theories at finite temperature and discussed under which conditions the link correlators evaluated in a given gauge could be used as order parameters for center symmetry. This analysis singles out certain gauges which we refer to as center symmetric.

We have then implemented one particular example of a center-symmetric gauge, which corresponds in the continuum to the center-symmetric Landau gauge of Ref.~\cite{vanEgmond:2021jyx}. Focusing on the link average, we have identified and solved the center symmetry constraints for this quantity. We have then tested these formal expectations on actual lattice data at temperatures below and above the transition. In the low temperature simulations, we find that the numerical link average accurately fulfills the center symmetry constraints, while the constraints are violated in the high temperature data. These numerical results shows very clearly that the link average plays indeed the role of an order parameter for center symmetry in that gauge. This result provides alternative ways to explore the confinement/deconfinement transition beyond the usual program based on the Polyakov loop. In frameworks, such as continuum calculations, where the Polyakov loop is not so easily accessed, this is certainly a valuable piece of information.

On the lattice side, a natural continuation of this work is of course the evaluation of the two-link correlator and the associated gluon propagator which has been shown to behave as an order parameter in this gauge as well, at least within the Curci-Ferrari set-up, see Ref.~\cite{vanEgmond:2023lfu}. The same goes for the ghost propagator. It would also be interesting to extend the analysis to the SU($2$) case where the transition is second order. In this case, we expect, if not a divergence, a large enhancement of the zero-momentum propagator at the transition, along the lines of what is observed in Ref.~\cite{vanEgmond:2022nuo}.

Other possible directions include the study of higher-order link correlators for which there are also symmetry constraints \cite{vanEgmond:2023lfu}, as well as the inclusion of dynamical quarks. In this latter case, center symmetry is broken explicitly by the quark boundary conditions and the transition becomes a crossover for small enough quark masses. Still, we expect the link average to grant access to the crossover temperature.

\begin{acknowledgments}
This work was supported by FCT - Fundaç\~ao para a Ciência e a Tecnologia, I.P., under Projects Nos. UIDB/04564/2020 \cite{1}, UIDP/04564/2020 \cite{2} and CERN/FIS-PAR/0023/2021 \cite{3}. P.~J.~S. acknowledges financial support from FCT contract CEECIND/00488/2017 \cite{4}. The authors acknowledge the Laboratory for Advanced Computing at the University of Coimbra (http://www.uc.pt/lca) and the  Minho Advanced Computing Center (http://macc.fccn.pt) for providing access to the HPC resources. Access to Navigator was partly supported by the FCT Advanced Computing Projects 2021.09759.CPCA \cite{5}, 2022.15892.CPCA.A2 and 2023.10947.CPCA.A2. Access to Deucalion was supported by the FCT Advanced Computing Project 2024.11063.CPCA.A3. P.~J.~S. would like to thank the financial support of CNRS and the hospitality of CPHT, École Polytechnique, where part of this work was performed. The authors would also like to welcome Olívia Maria who arrived in this world upon completion of this work. Godspeed!
\end{acknowledgments}

\appendix

\section{Inferring Eq.~(\ref{eq:csLandau}) from the continuum} \label{app:guess}
The continuum center-symmetric Landau gauges are particular instances of the family of background Landau gauges, defined by the condition
\beq
D[\bar A]_\mu(A_\mu-\bar A_\mu)=0\,,\label{eq:gfc}
\eeq
with $\bar A_\mu$ a background field configuration parametrizing the family of gauges  and $D[\bar A]_\mu\equiv\partial_\mu\mathds{1}+i[\bar A_\mu,\,\,]$ the adjoint covariant derivative for that background. For a generic background $\bar A_\mu$, the gauge fixing condition (\ref{eq:gfc}) is not invariant under center transformations, in the sense that, if $A_\mu$ is a gauge configuration obeying (\ref{eq:gfc}) for some background $\bar A_\mu$, then
\beq
D[\bar A^g]_\mu(A_\mu^g-\bar A_\mu^g)=gD[\bar A](A-\bar A)g^\dagger=0\,,
\eeq
which means that $A_\mu^g$ is a gauge configuration in a different background gauge, with background $\bar A_\mu^g$. 

\begin{widetext}
\begin{center}
\begin{figure}[t]
\includegraphics[height=0.16\textheight]{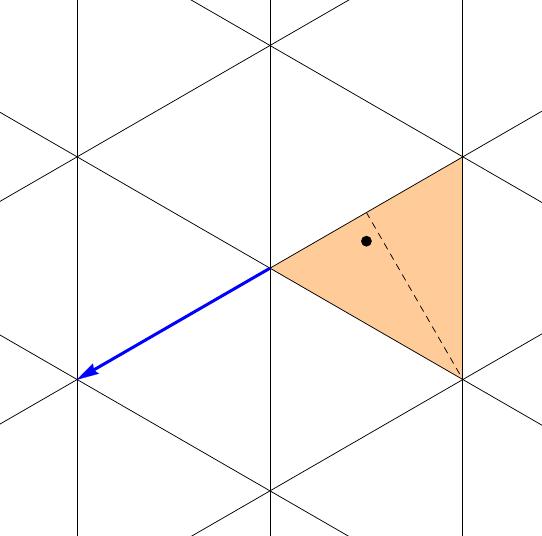}\qquad\includegraphics[height=0.16\textheight]{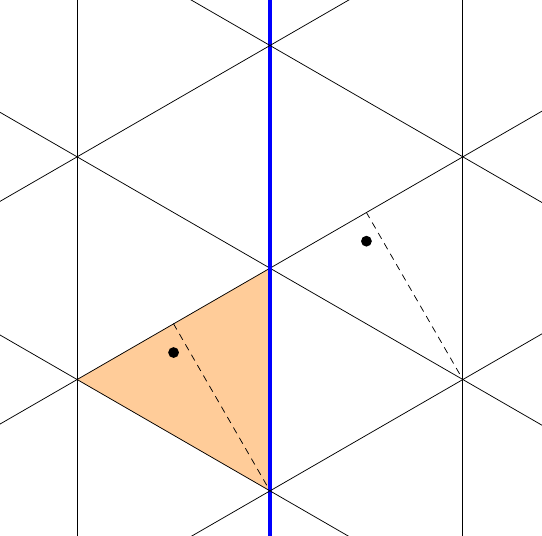}\qquad\includegraphics[height=0.16\textheight]{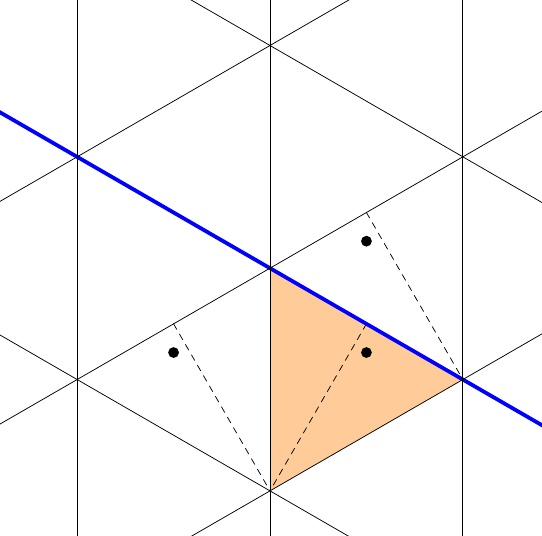}\qquad\includegraphics[height=0.16\textheight]{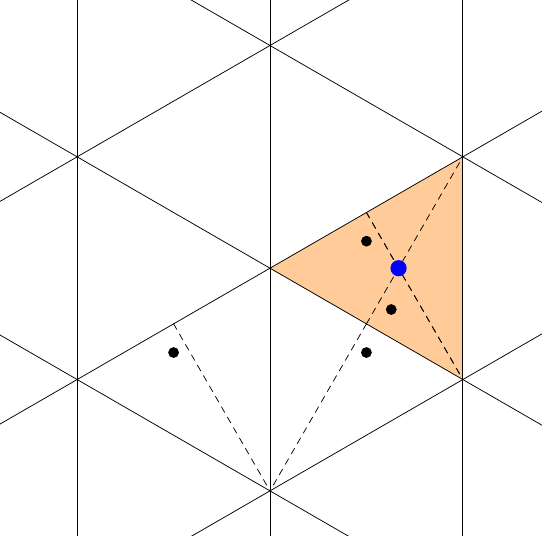}
\caption{Transformation of a Weyl chamber under a center transformation. The colored chamber represents the various locations of the Weyl chamber along the transformation process. We have chosen a point and a particular axis of the Weyl chamber to ease orientation as the Weyl chamber is transformed. In the first three figures, the blue items represent the transformations that will be applied to the Weyl chamber, $V_{-\rho_1}(\tau)$, $W_{\alpha_{12}}$ and $W_{\alpha_{31}}$ respectively, while in the fourth figure, the blue item represents the combined effect of these three transformations, which corresponds to a transformation of the original Weyl chamber into itself, more specifically a rotation by an angle $2\pi/3$ around its center.}
\label{fig:SU3_center}
\end{figure}
\end{center}
\end{widetext}

We emphasize that, in line with the lattice conventions chosen in the main text, the transformation of the gauge field is here
\beq
A_\mu^g(x)\equiv g(x)A_\mu(x) g^{\dagger}(x)-ig(x)\partial_\mu g^\dagger(x)\,.
\eeq
This sign is different from the convention used in Ref.~\cite{vanEgmond:2021jyx} but one can simply go from one convention to the other upon changing $A\to -A$. This also explains the different sign convention in the covariant derivative given above as compared to the one in Ref.~\cite{vanEgmond:2021jyx}. Note also that what we have here defined as the Polyakov loop would correspond to the anti-Polyakov loop in that reference.

\subsection{Center-symmetric Landau gauges}
Center-symmetric Landau gauges are obtained when choosing particular backgrounds $\bar A_{c,\mu}$, known as center-symmetric, which are invariant under particular center transformations $\tilde g$, that is 
\beq
\bar A_{c,\mu}^{\tilde g}=\bar A_{c,\mu}\,,\label{eq:Ac}
\eeq
In this case, $A_\mu$ and $A_\mu^{\tilde g}$ are both configurations in the same background gauge, of associated background $\bar A_{c,\mu}$.

One particularly simple way to construct center-symmetric backgrounds and the associated transformations $\tilde g$ is to restrict to constant, temporal and Abelian backgrounds:
\beq
\bar A_\mu(x)=T\delta_{\mu 0}\left(\bar r^3\frac{\lambda^3}{2}+\bar r^8\frac{\lambda^8}{2}\right).\label{eq:form}
\eeq 
In this subspace, one shows that the periodic gauge transformations are all generated by certain reflections in the plane $(\bar r^3,\bar r^8)$. The axes of these reflections form a lattice which subdivides the plane $(\bar r^3,\bar r^8)$ into regions known as Weyl chambers, all equivalent from a physical viewpoint since related to each other by periodic gauge transformations.

In this same subspace, the typical effect of a center transformation $g$ is to change a given Weyl chamber into a different one. However, using a certain number of the above mentioned reflections, corresponding to a certain periodic transformation $g_0$, one can always bring this particular Weyl chamber into its original position. This procedure then yields a particular center transformation $\tilde g$ that transforms this Weyl chamber into itself, and, thus, that leaves the barycenter of the Weyl chamber invariant. By construction, the background configuration corresponding to this barycenter obeys Eq.~(\ref{eq:Ac}). 

One example of center-symmetric background obtained using this procedure is
\beq
\bar A_{c,\mu}(x)=-T\delta_{\mu 0}\frac{4\pi}{3}\frac{\lambda^3}{2}\,,\label{eq:exc}
\eeq 
with the associated center transformation \cite{vanEgmond:2023lfu}
\beq
\tilde g(\tau)=W_{\alpha_{13}}W_{\alpha_{12}}e^{-i\frac{\tau}{\beta}4\pi\rho_1\cdot t}\,,
\eeq
corresponding to the combination of the transformations depicted Fig.~\ref{fig:SU3_center}. The different sign in Eq.~(\ref{eq:exc}) as compared to Ref.~\cite{vanEgmond:2023lfu} stems from the different convention used in the present work, as explained above.

\subsection{Connection to the Landau gauge}
Coming back to the family of background Landau gauges, let us now show that, in the case of a constant background, of which (\ref{eq:exc}) is a particular case, the gauge condition (\ref{eq:gfc}) over gauge fields $A_\mu$ that are periodic along the temporal direction, is equivalent to the standard Landau gauge condition but over gauge fields obeying different boundary conditions, see also Ref.~\cite{vanEgmond:2021jyx}. To this purpose, we consider a generic gauge transformation of both the gauge field and the background:
\beq
A_\mu^h & = & h A_\mu h^\dagger-ih\partial_\mu h^\dagger\,,\\
\bar A_\mu^h & = & h \bar A_\mu h^\dagger-ih\partial_\mu h^\dagger\,.
\eeq
By generic, we mean that there is nothing specific about $h$, not even its boundary conditions.

Exploiting the properties of the covariant derivative, we can write
\beq
D[\bar A]_\mu(A_\mu-\bar A_\mu)=h^\dagger\Big(D[\bar A^h]_\mu(A^h_\mu-\bar A^h_\mu)\Big)h\,.\label{eq:Ah}
\eeq
Consider now the particular case of
\beq
h=g_{\bar A}\equiv e^{i\tau\bar A}\,.\label{eq:gAb}
\eeq
Then
\beq
\bar A^{g_{\bar A}}=e^{i\tau\bar A}\bar A e^{-i\tau\bar A}-ie^{i\tau\bar A}(-i\bar A) e^{-i\tau\bar A}=0\,.
\eeq
In this case, Eq.~(\ref{eq:Ah}) becomes
\beq
D[\bar A]_\mu(A_\mu-\bar A_\mu)=g_{\bar A}^\dagger\Big(\partial_\mu A_\mu^{g_{\bar A}}\Big)g_{\bar A}\,,
\eeq
and, thus, for any gauge field configuration $A_\mu$ obeying (\ref{eq:gfc}), the particular  transformed gauge field $A_\mu^{g_{\bar A}}$ obeys the standard Landau gauge condition. 

The price to pay is of course that, in general, $A_\mu^{g_{\bar A}}$ is not anymore periodic along the temporal direction. This is because (\ref{eq:gAb}) is not periodic, not even modulo an element of the center. However, we can now exploit this equivalence between the background Landau gauges (for constant background) and the standard Landau gauge in order to infer the functional of the links that describes the background Landau gauge. Indeed, the functional of the links that describes the Landau gauge is ${\rm Re}\,\sum_{\mu,n} {\rm tr}\,U_\mu(n)$. We deduce that the functional of the links that describes a background Landau gauge with constant background $\bar A$ is simply
\beq
F_{\bar A}[U] & = & {\rm Re}\,\sum_{\mu,n} {\rm tr}\,U_\mu^{g_{\bar A}}(n)\nonumber\\
& = & {\rm Re}\,\sum_{\mu,n} {\rm tr}\,g_{\bar A}(n)\,U_\mu(n)\,g^\dagger_{\bar A}(n+\hat\mu)\,.
\eeq
Here, $g_{\bar A}(n)$ denotes the lattice version of (\ref{eq:gAb}):
\beq
g_{\bar A}(n)=e^{i\frac{n_4}{L_4}\beta\bar A}\,.
\eeq
Upon using the cyclicality of the trace, one finds
\beq
F_{\bar A}[U]={\rm Re}\,\sum_{\mu,n} {\rm tr}\,g^\dagger_{\bar A}(\hat\mu)\,U_\mu(n)\,,
\eeq
which, in the case of the center-symmetric background (\ref{eq:exc}) is nothing but (\ref{eq:csLandau}).\\


\section{Check of Eqs.~(\ref{eq:17}) and (\ref{eq:54})}\label{app:check}

We write
\begin{widetext}

\beq
\tilde g^\dagger(n+\hat4) g^\dagger_c(\hat4) \tilde g(n) & = & e^{i\frac{n_4+1}{L_4}\pi\left(\lambda_3+\frac{\lambda_8}{\sqrt{3}}\right)}e^{-i\pi\frac{\lambda_1}{2}}e^{-i\pi\frac{\lambda_4}{2}}e^{i\frac{4\pi}{3L_4} \frac{\lambda^3}{2}}e^{i\pi\frac{\lambda_4}{2}}e^{i\pi\frac{\lambda_1}{2}}e^{-i\frac{n_4}{L_4}\pi\left(\lambda_3+\frac{\lambda_8}{\sqrt{3}}\right)}\nonumber\\
& = & e^{i\frac{n_4+1}{L_4}\pi\left(\lambda_3+\frac{\lambda_8}{\sqrt{3}}\right)}e^{-i\pi\frac{\lambda_1}{2}}\left(
\begin{array}{ccc}
0 & 0 & -i\\
0 & 1 & 0\\
-i & 0 & 0 
\end{array}
\right)
\left(
\begin{array}{ccc}
e^{i\frac{2\pi}{3L_4}}  & 0 & 0\\
0 & e^{-i\frac{2\pi}{3L_4}} & 0\\
0 & 0 & 1
\end{array}
\right)
\left(
\begin{array}{ccc}
0 & 0 & i\\
0 & 1 & 0\\
i & 0 & 0 
\end{array}
\right)e^{i\pi\frac{\lambda_1}{2}}e^{-i\frac{n_4}{L_4}\pi\left(\lambda_3+\frac{\lambda_8}{\sqrt{3}}\right)}\nonumber\\
& = & e^{i\frac{n_4+1}{L_4}\pi\left(\lambda_3+\frac{\lambda_8}{\sqrt{3}}\right)}e^{-i\pi\frac{\lambda_1}{2}}\left(
\begin{array}{ccc}
1  & 0 & 0\\
0 & e^{-i\frac{2\pi}{3L_4}} & 0\\
0 & 0 & e^{i\frac{2\pi}{3L_4}}
\end{array}
\right)
e^{i\pi\frac{\lambda_1}{2}}e^{-i\frac{n_4}{L_4}\pi\left(\lambda_3+\frac{\lambda_8}{\sqrt{3}}\right)}\nonumber\\
& = & e^{i\frac{n_4+1}{L_4}\pi\left(\lambda_3+\frac{\lambda_8}{\sqrt{3}}\right)}\left(
\begin{array}{ccc}
0 & -i & 0\\
-i & 0 & 0\\
0 & 0 & 1 
\end{array}
\right)\left(
\begin{array}{ccc}
1  & 0 & 0\\
0 & e^{-i\frac{2\pi}{3L_4}} & 0\\
0 & 0 & e^{i\frac{2\pi}{3L_4}}
\end{array}
\right)
\left(
\begin{array}{ccc}
0 & i & 0\\
i & 0 & 0\\
0 & 0 & 1 
\end{array}
\right)e^{-i\frac{n_4}{L_4}\pi\left(\lambda_3+\frac{\lambda_8}{\sqrt{3}}\right)}\nonumber\\
& = & e^{i\frac{n_4+1}{L_4}\pi\left(\lambda_3+\frac{\lambda_8}{\sqrt{3}}\right)}\left(
\begin{array}{ccc}
e^{-i\frac{2\pi}{3L_4}}  & 0 & 0\\
0 & 1 & 0\\
0 & 0 & e^{i\frac{2\pi}{3L_4}}
\end{array}
\right)e^{-i\frac{n_4}{L_4}\pi\left(\lambda_3+\frac{\lambda_8}{\sqrt{3}}\right)}\nonumber\\
& = & e^{i\frac{n_4+1}{L_4}\pi\left(\lambda_3+\frac{\lambda_8}{\sqrt{3}}\right)}e^{-i\frac{\pi}{L_4}\left(\frac{\lambda_3}{3}+\frac{\lambda_8}{\sqrt{3}}\right)}e^{-i\frac{n_4}{L_4}\pi\left(\lambda_3+\frac{\lambda_8}{\sqrt{3}}\right)}\nonumber\\
& = & e^{i\frac{\pi}{L_4}\left(\lambda_3+\frac{\lambda_8}{\sqrt{3}}\right)}e^{-i\frac{\pi}{L_4}\left(\frac{\lambda_3}{3}+\frac{\lambda_8}{\sqrt{3}}\right)}=e^{i\frac{2\pi}{3L_4}\lambda_3}=e^{i\frac{4\pi}{3L_4}\frac{\lambda_3}{2}}=g^\dagger_c(\hat 4)\,.
\eeq
Similarly,
\beq
& & e^{i\pi\frac{\lambda_1}{2}}g_c(\hat\mu) e^{-i\pi\frac{\lambda_1}{2}}\nonumber\\
& & \hspace{0.5cm}=\,\left(
\begin{array}{ccc}
0 & -i & 0\\
-i & 0 & 0\\
0 & 0 & 1 
\end{array}
\right)\left(
\begin{array}{ccc}
e^{-i\frac{2\pi}{3L_4}}  & 0 & 0\\
0 & e^{i\frac{2\pi}{3L_4}} & 0\\
0 & 0 & 1
\end{array}
\right)
\left(
\begin{array}{ccc}
0 & i & 0\\
i & 0 & 0\\
0 & 0 & 1 
\end{array}
\right)=\,g^\dagger_c(\hat\mu)\,.\label{eq:CC}
\eeq

\section{Link averages}\label{app:averages}
The more precise form of the link averages in Sec.~\ref{SecResults} are
\begin{eqnarray}
& & 
\langle U_4 \rangle = 
~ \left(
\begin{array}{ccc}
0.8687(92)   -0.123(26) \, i &      0.023(27)   +0.014(19)  \, i &   
      0.020(18)   -0.049(19) \, i \\
   -0.001(27)    +0.015(20)  \, i &      0.8662(99)  +0.131(26)  \, i &  -0.001(20)   +0.004(21) \, i \\
   -0.027(18)    -0.040(20)  \, i &     -0.002(22)   +0.006(20)  \, i &   0.8853(86)  -0.009(24) \, i
\end{array}
\right),
\\
& &
\langle U^{\tilde g}_4 \rangle = 
~ \left(
\begin{array}{ccc}
0.829(12)  -0.312(21)  \, i &     0.005(18)   -0.048(21) \, i &   0.006(21)   -0.000(23) \, i \\
   0.002(18)  -0.0522(21) \, i &     0.744(16)   +0.464(21) \, i &   0.005(21)   -0.026(25) \, i \\
  -0.004(25)  +0.000(19)  \, i &    -0.0110(25)  -0.011(22) \, i &   0.8588(86)  -0.173(23) \, i
\end{array}
\right),
\\
& &
\langle U^{\tilde{g}^\dagger}_4 \rangle = 
~ \left(
\begin{array}{ccc}
0.748(14)   -0.457(22) \, i  &    -0.003(21)7  -0.002(21)  \, i &   -0.00(25)    +0.014(24) \, i \\
    0.006(19)   -0.002(26) \, i  &     0.8352(97)  +0.294(24)  \, i &   -0.047(19)   +0.012(18) \, i \\
    0.026(25)   -0.004(22) \, i  &     0.039(18)   +0.035(20)  \, i &    0.858(11)   +0.182(25) \, i
\end{array}
\right),
\end{eqnarray}
for the ensemble averages of the temporal link at the origin of the high temperature $64^3 \times 6$ lattice, and
\beq
& & 
\langle U_4 \rangle = 
~ \left(
\begin{array}{ccc}
0.829(10)  -0.194(25) \, i &   -0.021(21)   +0.054(24) \, i  &   -0.022(24)   +0.013(26) \, i \\
    0.006(23)  +0.045(24) \, i &    0.827(11)   +0.241(28) \, i  &    0.003(24)   -0.022(19) \, i \\
    0.024(24)  +0.024(24) \, i &    0.018(24)   -0.020(21) \, i  &    0.8476(95)  -0.050(27) \, i
\end{array}
\right),
\\
& &
\langle U^{\tilde g}_4 \rangle = 
~ \left(
\begin{array}{ccc}
   0.806(12)   -0.267(28)  \, i &  0.009(23)7  +0.033(22)  \, i &  -0.024(21)   -0.012(26) \, i \\
   -0.017(24)   +0.018(24)  \, i &  0.815(11)   +0.247(25)  \, i &   0.058(22)   +0.006(24) \, i \\
    0.02(21) 2  -0.003(25)  \, i & -0.042(22)   -0.017(25)  \, i &   0.8611(99)  +0.019(26) \, i
\end{array}
\right),
\\
& &
\langle U^{\tilde{g}^\dagger}_4 \rangle = 
~ \left(
\begin{array}{ccc}
 0.837(13)   -0.204(26)5 \, i &    0.020(21)2  +0.009(24)  \, i &   -0.006(25)   +0.045(23) \, i \\
   -0.026(20)   -0.006(27)  \, i &    0.832(11)   +0.171(27)  \, i &    0.029(25)   -0.017(22) \, i \\
    0.009(24)   +0.057(22)  \, i &   -0.007(27)   -0.024(23)  \, i &    0.8506(97)  +0.027(26) \, i
\end{array}
\right),
\eeq
for the ensemble averages of the temporal link at the origin of the low temperature $64^3 \times 8$ lattice.

As for the combined ensemble and lattice averages, we find
\beq
& & 
\langle\langle U_4 \rangle\rangle = \nonumber \\
& & 
~ \left(
\begin{array}{ccc}
0.853201(34)  -0.14567(16)   \, i &    -0.0000002(71) +0.0000078(69)   \, i &   -0.0000038(70)  -0.0000023(72) \, i \\
   -0.0000002(71) -0.0000078(69) \, i &     0.853206(40)  +0.14559(18)     \, i &   -0.0000094(81)  -0.0000087(70) \, i \\  
   -0.0000028(68) +0.0000035(72) \, i &    -0.0000118(80) +0.0000050(72)   \, i &    0.866012(17)   +0.00009(15) \, i
\end{array}
\right),\nonumber\\
\\
& & 
\langle\langle U^{\tilde g}_4 \rangle\rangle = \nonumber \\
& & 
~ \left(
\begin{array}{ccc}
0.813815(56)  -0.29611(14)   \, i &    0.0000140(56)  -0.0000043(65) \, i &     0.0000087(72)  +0.0000094(83) \, i \\
       0.0000149(56) +0.0000043(65) \, i &    0.747226(90)   +0.43683(16)   \, i &     0.0000056(61)  -0.0000015(76) \, i \\
       0.0000050(75) -0.0000118(82) \, i &    0.0000048(59)  +0.0000034(75) \, i &     0.851544(32)   -0.15501(17) \, i
\end{array}
\right),\nonumber\\
\\
& & 
\langle\langle U^{\tilde{g}^\dagger}_4 \rangle\rangle = \nonumber \\
& & 
~ \left(
\begin{array}{ccc}
0.747174(82)  -0.43691(15)   \, i &    -0.0000050(66)  -0.0000038(59) \, i &     0.0000015(75)  -0.0000056(65) \, i \\
      -0.0000050(66) +0.0000038(59) \, i &     0.813756(52)1  +0.29627(13)   \, i &     0.0000023(75)  +0.0000038(73) \, i \\
       0.0000033(74) +0.0000048(62) \, i &     0.0000035(71)  -0.0000028(73) \, i &     0.851569(32)   +0.15492(16)    \, i 
\end{array}
\right),\nonumber\\
\eeq
over the high temperature $64^3 \times 6$ lattice, and
\beq
& & 
\langle\langle U_4 \rangle\rangle = \nonumber \\
& & 
~ \left(
\begin{array}{ccc}
0.83362(15)   -0.21910(51)   \, i &   -0.0000013(84) -0.0000034(86) \, i &   -0.000001(10)   +0.0000084(94) \, i \\
  -0.0000013(84) +0.0000034(86) \, i &    0.83352(12)   +0.21939(45) \,   i &   -0.0000008(85)  -0.0000084(82) \, i \\
  -0.000003(10)  -0.0000079(88) \, i &   -0.0000029(81) +0.0000079(83) \, i &    0.861665(24)   -0.00031(50) \, i
\end{array}
\right),\nonumber\\
\\
& & 
\langle\langle U^{\tilde g}_4 \rangle\rangle = \nonumber \\
& & 
~ \left(
\begin{array}{ccc}
0.83222(14)   -0.22331(53)   \, i &  0.0000109(81) +0.0000101(86) \, i &   0.0000084(85) +0.0000008(86) \, i \\
       0.0000109(81) -0.0000101(86) \, i &   0.83149(14) +0.22706(52) \, i &  -0.0000024(78) -0.0000097(90) \, i \\
       0.0000079(84) -0.0000029(80) \, i &  -0.0000048(82) +0.0000087(85) \, i &   0.861905(26)  -0.00382(46) \, i
\end{array}
\right),\nonumber\\
\\
& & 
\langle\langle U^{\tilde{g}^\dagger}_4 \rangle\rangle = \nonumber \\
& & 
~ \left(
\begin{array}{ccc}
0.83155(12)   -0.22676(42)   \, i &    0.0000010(88) -0.0000072(86) \, i &   0.0000097(91) +0.0000024(79) \, i \\
       0.0000010(88)  0.0000072(86) \, i &    0.83238(14)   +0.22272(51)   \, i &  -0.0000084(90) +0.0000009(93) \, i \\
       0.0000087(88) -0.0000048(83) \, i &   -0.0000079(89) -0.0000031(95) \, i &   0.861921(23)  +0.00412(49) \, i 
\end{array}
\right),\nonumber\\
\eeq
over the low temperature $64^3 \times 8$ lattice.

\end{widetext}

\pagebreak

\section{Basic steepest ascent algorithm}\label{app:ascent}

Let us here show that the basic steepest ascent algorithm applied to a center-symmetric gauge functional $\tilde F[U]$ is symmetry preserving in the sense that, starting from two configurations $U$ and $U^{\tilde g}$ related by the transformation $\tilde g$ that leaves the functional $\tilde F[U]$ invariant, the two maxima $U^{g_0}$ and $(U^{\tilde g})^{g_0'}$ that are selected by the steepest ascent algorithm, along the ${\cal G}_0$-orbits of $U$ and $U^{\tilde g}$, are also related by $\tilde g$:
\beq
(U^{\tilde g})^{g_0'}=(U^{g_0})^{\tilde g}\,.
\eeq

Let us first explain how a local maximum is found starting from $U_\mu(n)$. One first constructs the gradient
\beq
X^a[U](n)\equiv\left.\frac{\delta \tilde F[U^{e^{i\theta^bt^b}}]}{\delta\theta^a(n)}\right|_{\theta=0}\,,
\eeq
and builds an ascent gauge transformation
\beq
g_0[U;\lambda]=\exp\left\{i\lambda X^a[U]t^a\right\},
\eeq
with $\lambda$ some adjustable parameter, taking for instance the trial values $1\to1/2\to 1/4\to\cdots$ until $\tilde F[U^{g_0[U;\lambda]}]$ exceeds $\tilde F[U]$. After one step, the procedure produces one value of $\lambda$ denoted $\lambda_1$ as well as $U_1\equiv U^{g_0[U;\lambda_1]}$ which can be used as a new link to continue the ascent. Repeating the procedure many times, one obtains a collection of numbers $\lambda_i$ as well as intermediate links $U_i\equiv U^{g_0[U_{i-1};\lambda_i]}$  which are closer and closer to a maximum. The procedure stops whenever the gradient becomes compatible with zero in which case one has reached a maximum.\footnote{In fact it could happen that one reaches more generally an extremum. But looking for maxima is just a numerical convenience and, if one ends up with some extrema, there is no problem in including them in the definition of the gauge-fixed configurations.}

The question is now: what happens if we start from $U_\mu^{\tilde g}$ instead? How are the corresponding $\lambda_i$'s and intermediate links related to the previous ones? To answer this question, we use Eq.~(\ref{eq:Fc}) to write
\beq
\tilde F[(U^{\tilde g})^{g_0}] & = & \tilde F[U^{g_0\tilde g}]\nonumber\\
& = & \tilde F[(U^{\tilde g^\dagger g_0 \tilde g})^{\tilde g}]\nonumber\\
& = & \tilde F[U^{\tilde g^\dagger g_0 \tilde g}]\,.\label{eq:44}
\eeq 
From this, we can deduce a relation between the gradients at $U$ and $U^{\tilde g}$, namely (pay attention to the ordering of $\tilde g$, $t^a$ and $\tilde g^\dagger$ when performing the chain rule)
\beq
X^a[U^{\tilde g}](n)\,t^a=X^a[U](n)\,\tilde g(n)\,t^a \tilde g^\dagger(n)\,,
\eeq
from which it follows that
\beq
g_0[U^{\tilde g};\lambda]=\tilde g\,g_0[U,\lambda]\tilde g^\dagger\,,
\eeq
and thus
\beq
(U^{\tilde g})^{g_0[U^{\tilde g};\lambda]} & = & (U^{\tilde g})^{\tilde gg_0[U,\lambda]\tilde g^\dagger}\nonumber\\
& = &  U^{\tilde gg_0[U,\lambda]}\nonumber\\
& = &  (U^{g_0[U,\lambda]})^{\tilde g}\,.
\eeq
Then
\beq
\tilde F[(U^{\tilde g})^{g_0[U^{\tilde g};\lambda]}]=\tilde F[U^{g_0[U;\lambda]}]\,.
\eeq
This means the adjustment $\lambda_1$ is the same as before (if one uses the same algorithm $1\to 1/2\to 1/4\to \dots$). And moreover
\beq
V_1\equiv (U^{\tilde g})^{g_0[U^{\tilde g};\lambda_1]} & = & (U^{g_0[U,\lambda_1]})^{\tilde g}=U_1^{\tilde g}\,.
\eeq
\begin{widetext}

\begin{center}
\begin{figure}[t]
\includegraphics[height=0.14\textheight]{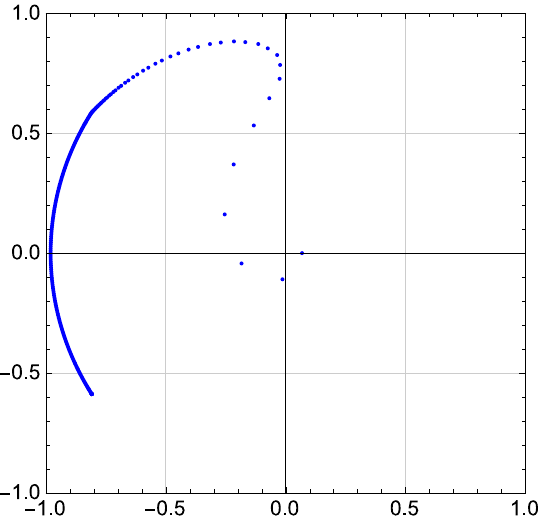}\qquad\includegraphics[height=0.14\textheight]{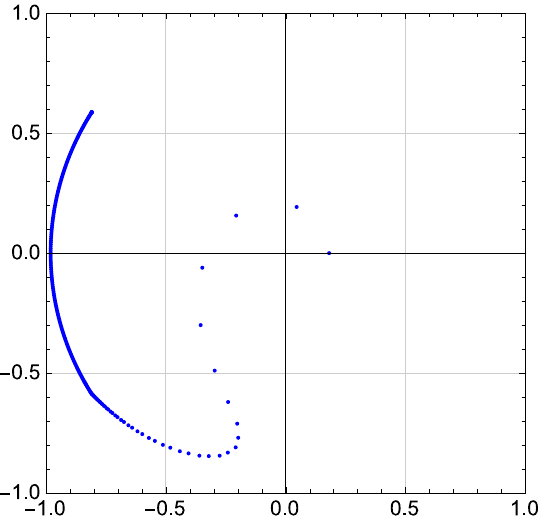}\qquad\includegraphics[height=0.14\textheight]{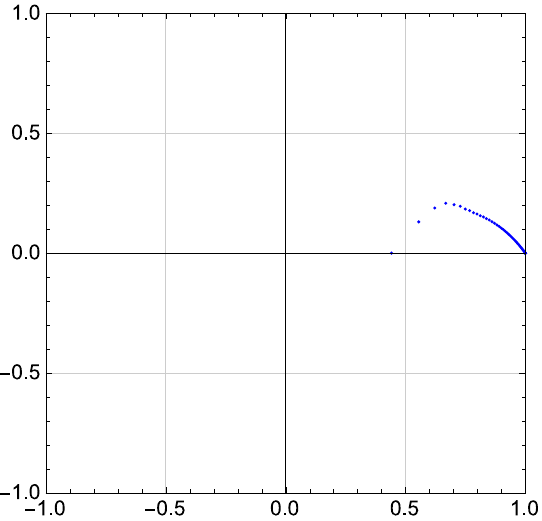}\qquad\includegraphics[height=0.14\textheight]{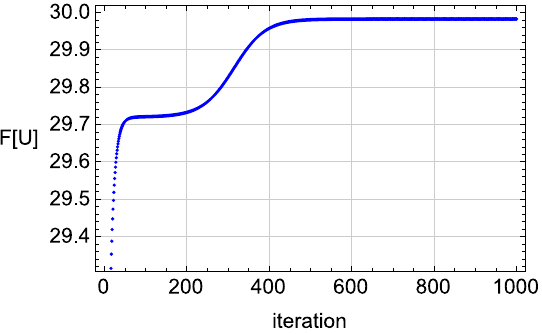}\\
\vglue4mm
\includegraphics[height=0.14\textheight]{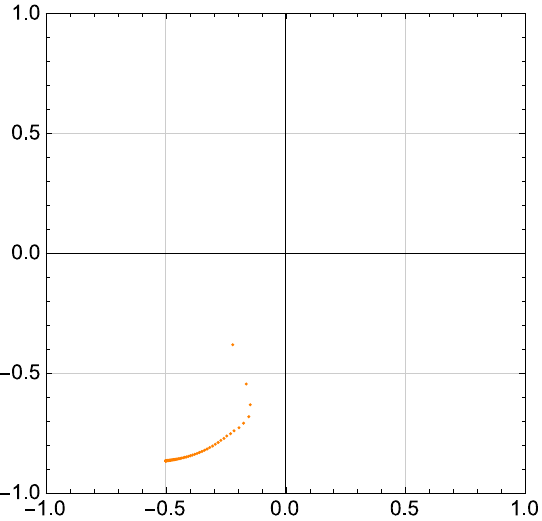}\qquad\includegraphics[height=0.14\textheight]{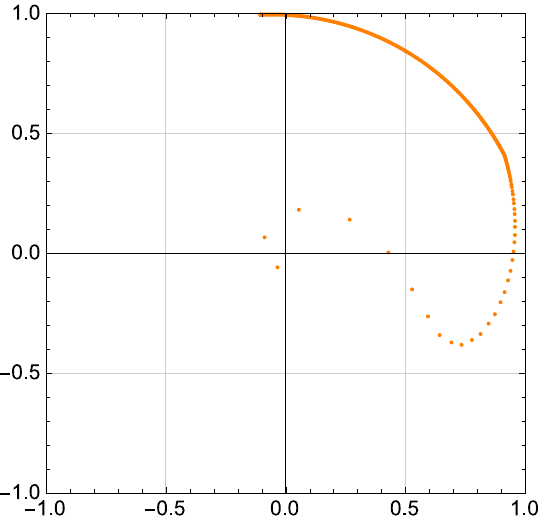}\qquad\includegraphics[height=0.14\textheight]{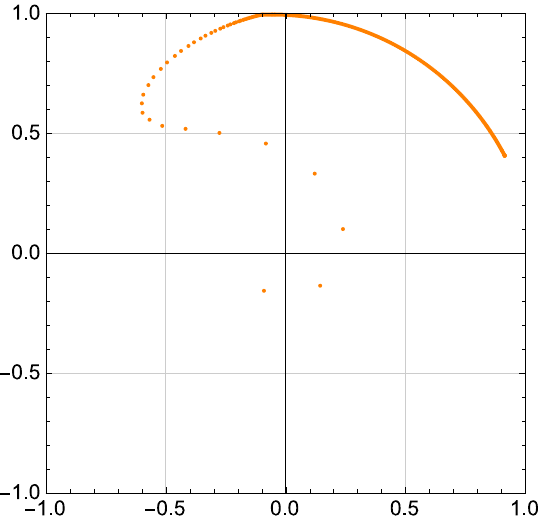}\qquad\includegraphics[height=0.14\textheight]{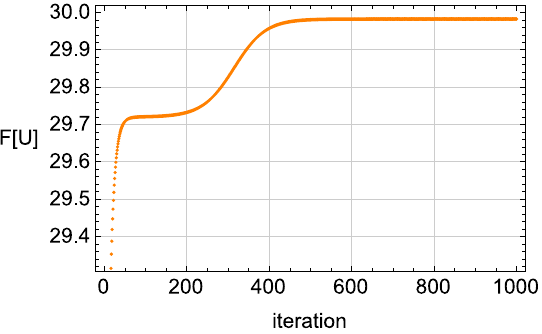}
\caption{The three diagonal components (from left to right but one) of the link elevated to the power $L$ (number of lattice sites) at a given site of a $1d$ lattice along the ascent to the maximum of $\tilde F[U]$. The top and bottom plots differ by the starting links which are chosen to be related by the center transformations $\tilde g$. The right-most plot shows the value of $\tilde F[U]$ along the ascent. This function is exactly the same for the two ascents considered here.}
\label{fig:VW}
\end{figure}
\end{center}

\end{widetext}
From this we deduce that, along the ascent from $U$ or $U^{\tilde g}$, all $\lambda_i$'s are the same and the intermediate links are related as $V_i=U_i^{\tilde g}$. This ensures that the final, extremal configurations are also related by $\tilde g$ which ensures that $\smash{{\cal E}_{\rm gf}^{\tilde g}\simeq {\cal E}_{\rm gf}}$ in the symmetric phase.

We have implemented the simple steepest ascent algorithm described in Sec.~\ref{sec:Gribov} for the functional (\ref{eq:csLandau}) on a one-dimensional lattice of $L$ sites. In Fig.~\ref{fig:VW}, we show, in the complex plane, the three diagonal components of the links, elevated to the power $L$ for convenience, along the ascent to the maximum of $\tilde F[U]$. We start from two different link configuration related by the transformation $\tilde g$ in Eq.~(\ref{eq:g}). The diagonal components of these initial configurations are related by (\ref{eq:69})-(\ref{eq:71}). By elevating the components to the power $L$, the relations involve simple phases $e^{\pm i2\pi/3}$ which one can easily identify in the plots. What the plots show is that these relations are preserved along the ascent to the maximum, thus confirming that this simple implementation of the steepest ascent algorithm preserves the $\tilde g$-invariance of the ensemble.

\end{document}